\documentclass[a4paper,11pt,pdftex,dvipsnames]{article}

\usepackage[utf8]{inputenc} 
\usepackage[T1]{fontenc}    
\usepackage{amssymb,amsmath,latexsym,amsfonts,amsthm}
\usepackage{hyperref}       
\usepackage{cleveref}
\usepackage{url}            
\usepackage{booktabs}       
\usepackage{amsfonts}       
\usepackage{nicefrac}       
\usepackage{microtype}      
\usepackage{graphicx}
\usepackage{doi}
\usepackage{tikz}
\usepackage{pgfplots}
\pgfplotsset{compat=1.18}
\usetikzlibrary{
  arrows,
  arrows.meta
}
\usepackage[
  autocite    = superscript,
  backend     = bibtex,
  sortcites   = true,
  style       = numeric,
  doi=false, url=true, isbn=false,
  maxcitenames=1, maxbibnames=1
  ]{biblatex}

\bibliography{references}

\usepackage{amsmath}
\usepackage{amssymb}

\usepackage[a4paper, total={6in, 8in}]{geometry}

\usepackage[OT2,T1]{fontenc}
\usepackage[OT2,T1]{fontenc}
\usepackage{verbatim}
\DeclareSymbolFont{cyrletters}{OT2}{wncyr}{m}{n}
\DeclareMathSymbol{\Sha}{\mathalpha}{cyrletters}{"58}
\usepackage{graphicx}
\usepackage{subcaption}
\newtheorem{theorem}{Theorem}
\numberwithin{equation}{section}

\graphicspath{ {./images/} }

\usepackage{authblk}

\title{Elastic waves in bearing raceways: the forward and inverse problem}

\author{Jessica J.\ Kent, Matheus de C.\ Loures, Art L.\ Gower${}^1$}
\affil{Department of Mechanical, Aerospace and Civil Engineering, The University of Sheffield, UK}

\newcommand{\lap}{\nabla^2}
\newcommand{\besselj}{\mathrm{J}}
\newcommand{\hankelh}{\mathrm{H}}
\newcommand{\e}{\mathrm{e}}

\newcommand{\im}{\mathrm{i}}
\newcommand{\ii}{\mathrm{i}}
\newcommand{\ee}{\mathrm{e}}

\newcommand{\inv}{\mathrm{inv}}
\newcommand{\for}{\mathrm{for}}
\newcommand{\Efor}{\mathbf E^\for}
\newcommand{\Mfor}{\mathbf M^\for}
\newcommand{\Minv}{\mathbf M^\inv}
\newcommand{\Einv}{\mathbf E^\inv}

\renewcommand{\vec}[1]{ \boldsymbol{#1}}

\usepackage[textsize=normalsize,textwidth=2cm,colorinlistoftodos]{todonotes}

\newcounter{todocounter}

\def\p{0.03}
\def\r{0.25}
\tikzset{
  wavefront/.pic={
    \tikzset{/wavefront/.cd,#1}
    \fill (0,0) circle (\p);
    \draw (\wang:\r) arc(\wang:-\wang:\r);
  }
  /wavefront/.search also={/tikz},
  /wavefront/.cd,
  ang/.store in=\wang, ang={60},
}

\begin{document}
\maketitle

\begin{abstract}
 Turbines are crucial to our energy infrastructure, and ensuring their bearings function with minimal friction while often supporting heavy loads is vital. Vibrations within a bearing can signal the presence of defects, friction, or misalignment. However, current detection methods are neither robust nor easy to automate. We propose a more quantitative approach by modelling the elastic waves within bearing raceways. By approximating the raceway as a hollow cylinder, we derive straightforward 4x4 systems for its vibrational modes, enabling both forward and inverse problem-solving. We also demonstrate how to significantly reduce the number of required sensors by using a simple prior: the known number of rollers and their angular speed. We present numerical examples showcasing the full recovery of contact traction between bearings and the raceway, as well as the detection of elastic emissions.
\end{abstract}

\section{Introduction}
\label{sec:intro}
Bearings are essential parts of modern industrial machinery; found everywhere from bicycles to wind turbines to jet engines \cite{harris2001rolling}. Their main purpose is to reduce friction and constrain the motion of rotating components; as such, their maintenance and efficiency remains an important industrial problem and an active field of study \cite{hart2019}.

\footnotetext[1]{Corresponding author: arturgower@gmail.com}

\noindent \textbf{Current methods.} The most successful methods to monitor the condition of roller bearings are based on vibration analysis: analysing the frequency components of how the raceway, or mounting, vibrates in time
\cite{randall2021vibration,Sinha2020industrial,rezaei2007fault}. 
Current methods perform best for identifying localised defects that create an impulsive response; given the right conditions these methods can identify whether there is a localised defect on the inner or outer race, or on the rollers themselves. 
However, it has been challenging to make current methods robust and reliable enough to automate, and therefore experts are often required to both process and interpret the results. 


\noindent \textbf{Too few sensors.} It is likely that robust diagnostics are not possible for a small number of sensors. Typically, sensors measure displacement or acceleration at one or two positions, and rotation speed on each large bearing \cite{lees2020vibration}. That is, there are often too many unknowns for the number of sensors typically used. Further, acceleration sensors are often placed on the housing, in which case the \emph{transfer path} of the signal can be unknown and complex. We call the \emph{transfer path} the route an elastic wave takes from its source to a sensor. 

\noindent \textbf{Be more quantitative.} In this work we suggest that to have more robust and automatic predictions we need to properly model the physics of elastic waves in the bearing raceway. After all, these elastics waves are what carry the information of the forces inside the bearing to the sensors. Modelling elastic waves enable us to: 1) determine how well we can truly expect a few sensors to perform, and 2) develop more quantitative methods to predict defects and contact forces in bearings. 

A typical approach to account for the physics of elastic waves is to use finite element methods, however these are far too computationally intensive and opaque for inverse problems \cite{singh2014analyses}. In this paper we show how elastic waves can be very efficiently described both for sensing and modelling, with details in the next section. 

\noindent \textbf{Nonlinear dynamics at interfaces.} For roller and journal bearings the way that forces are transmitted through the bearing and the shaft at interfaces are non-linear \cite{Champneys_2021, shaw2019normal, Janjarasjitt2008}. 
For example, a roller rattling is a non-linear dynamic event. Non-linearities at the interfaces make it challenging to accurately solve for the dynamics of an entire system, which 
typically includes components like shafts, rotors, bearings, and varying foundations \cite{Lees2009model}. Over the past 40 years there has been significant work to  model the entire system \cite{mcfadden1984model, mcfadden1985vibration,singh2015extensive,morais2020rotating,,lees2020vibration}.




\begin{figure}[h!]
    \centering
    \includegraphics[width=0.39\linewidth]{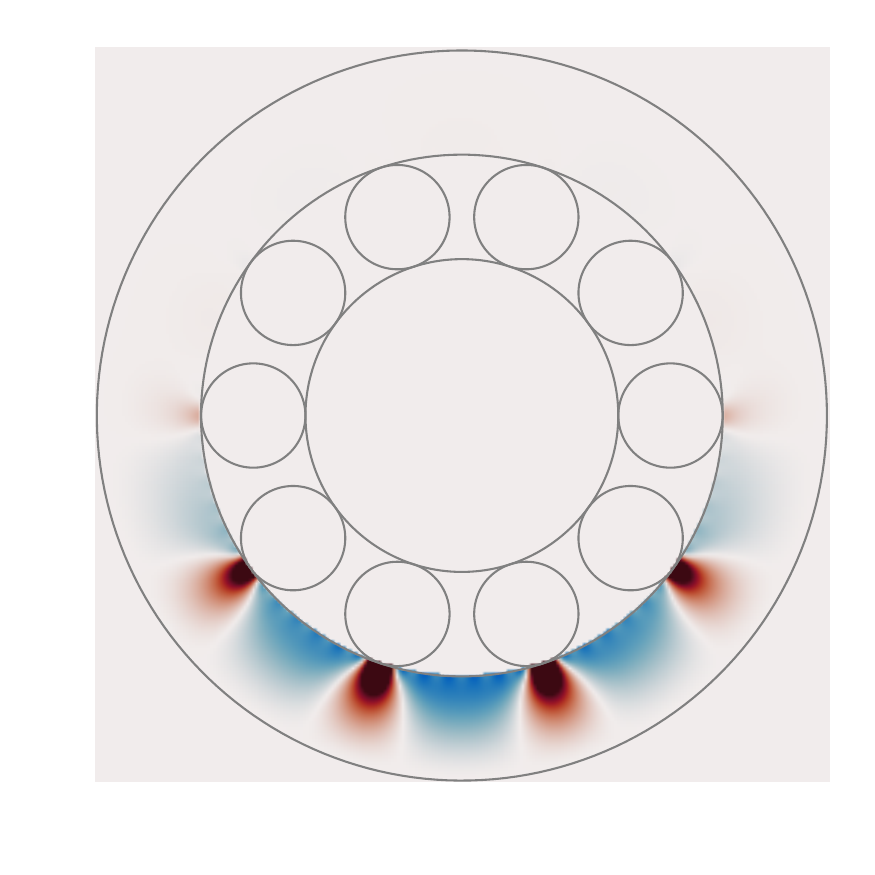}
    \includegraphics[width=0.39\linewidth]{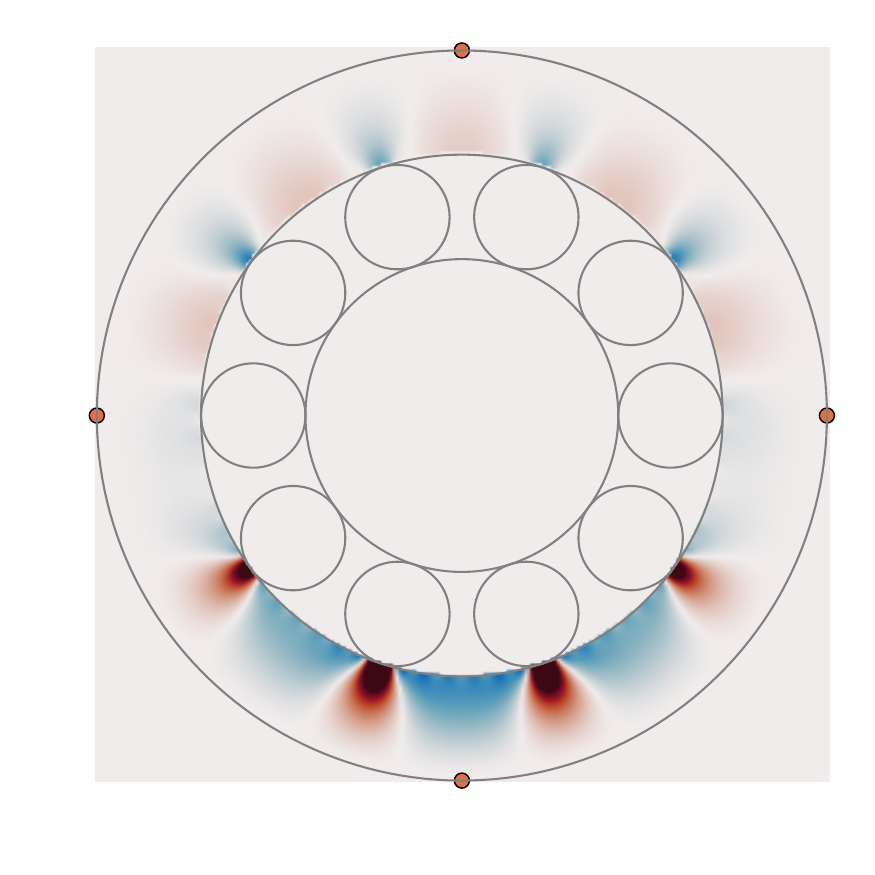}
    \caption{True pressure on the left for one snapshot in time. The right shows the predicted stress for the same time when using 4 sensors shown as orange spots.  The sensors measure displacement, and the outer boundary is stress free, which is why the stress tends to zero when it reaches the outer boundary. However, the displacements or accelerations caused by the stress are not zero on this boundary and can be measured. The recovery is not perfect as there is a 20\% error added to the boundary data. More details given in \Cref{sec:example-loading-profile}. }
    \label{fig:true and prior stress}
\end{figure}

\noindent \textbf{Linear elastic waves.} While the interface conditions between machine components are often non-linear, the elastic waves within each component are primarily governed by linear elasticity.
This allows us to break the problem into manageable parts: by measuring the vibrations at one boundary, we can predict the forces or vibrations at another boundary of the same component. 

We demonstrate that it is possible to accurately predict the stresses on a bearing raceway. See \Cref{fig:true and prior stress} for a motivational example which predicts the stresses between the rollers and the raceway with just four sensors. There is currently no such method to predict the forces in a bearing which are important for lifetime analysis and to understand the causes of defects \cite{howard94}.  


\noindent \textbf{Acoustic emission and localisation.} One way to diagnose when defects appear is to measure the sound a defect emits when it forms or expands \cite{eaton2012towards, jones2022bayesian,jones2021bayesian}. These methods rely on measuring only pressure waves in solid components and are almost exclusively based in the time domain, where the first signal that arrives is (likely) the bulk pressure mode. It is important to only use the first measured signal, as the next signals will be a mix of acoustic and shear waves that have likely mode converted at boundaries, and, especially for thin structures, potentially formed waveguide modes. To extract the first signal can be difficult when its amplitude is far less than the other wave modes, and, of course, its arrival time is unknown. \cite{grosse2021acoustic} 

Alternatively, instead of assuming that there are only acoustics waves, one could just measure the signal in time, potentially both pressure and shear displacement, and from that determine where the source is. To do so would require modelling elastic waves. This paper is the first step to an elastic emission method in bearing raceways that does not require extracting the first arrival time, and could continually measure and identify sources. 



\noindent \textbf{Paper contents.} In this paper we show theoretical and numerical results on how to predict the stress on the raceway and between rollers and the raceway. The methods developed can be specialised to other bearings, though we focus on roller bearings here. 

In \Cref{sec:maths} we discuss how, at high enough frequencies, elastic waves are mostly confined to the bearing raceway; this allows us to model the waves in a raceway as vibrations of a hollow cylinder, as opposed to modelling the full bearing system. In \Cref{sec:bcs} we  develop a modal method and show how to use boundary conditions to quickly calculate elastic waves in the raceway. 

In \Cref{sec:prior method} we show how assumptions about the boundary conditions, which we call priors, can greatly reduce the number of sensors required. As an example, in \Cref{sec:roller bearings} we show how to deduce and use priors about rollers rotating at a constant speed.

From the elastic wave models we learn what is, and is not, measurable, which we summarise in \Cref{sec:what is measureable}. In \Cref{sec:examples} we show several numerical examples both for validation and to illustrate the main results.


\section{Elastic waves in raceways}
\label{sec:maths}
\begin{figure}[h]
    \centering
    \includegraphics[width=0.5\linewidth]{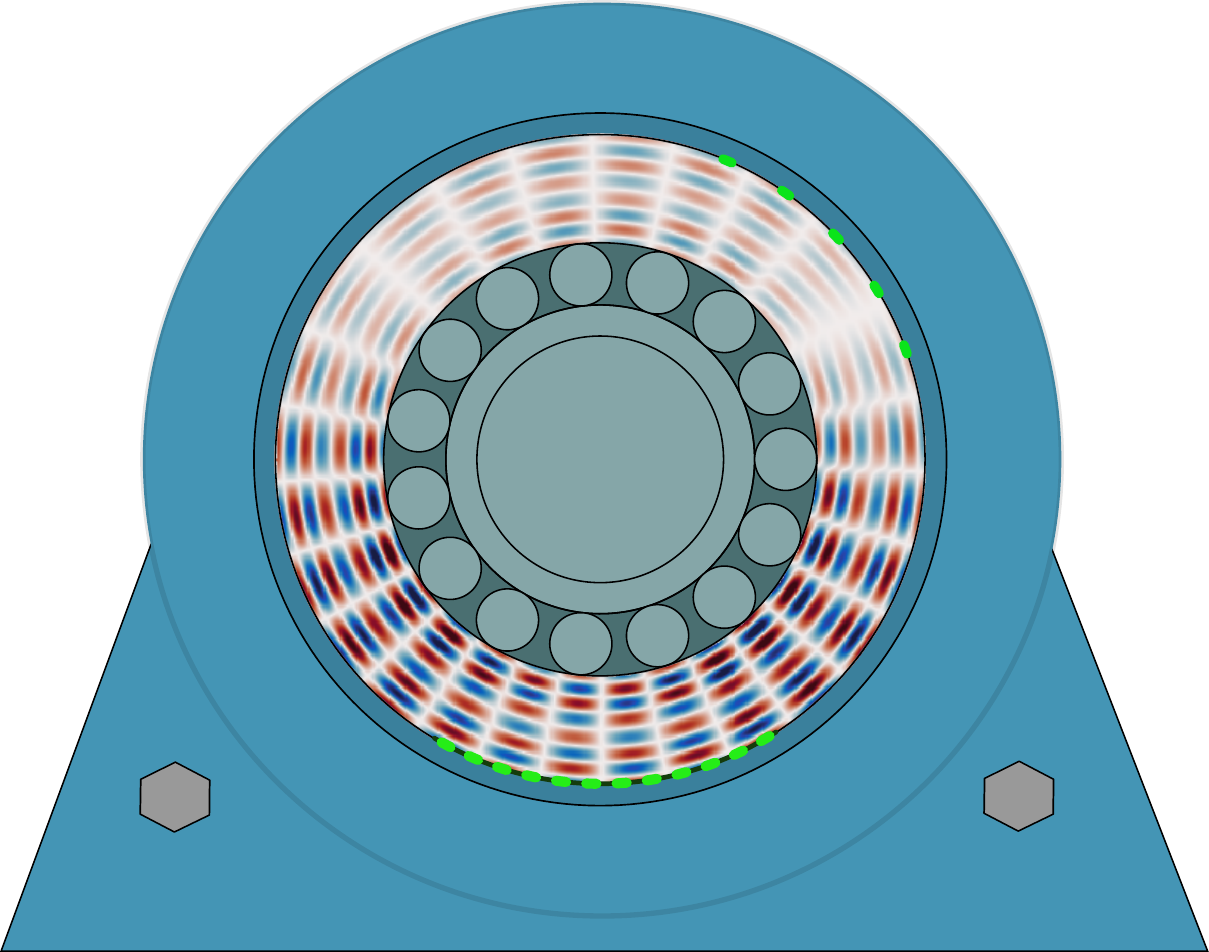}
    \caption{A cross section illustration of a bearing in a mounting. Showing a wave for a fixed frequency (> 10kHzs) showing waves trapped in the raceway. Further shown are sensors mounted on the raceway as we propose.}
    \label{fig:raceway in mounting}
\end{figure}

\begin{figure}[ht]
    \centering
  \includegraphics[width=0.3\linewidth]{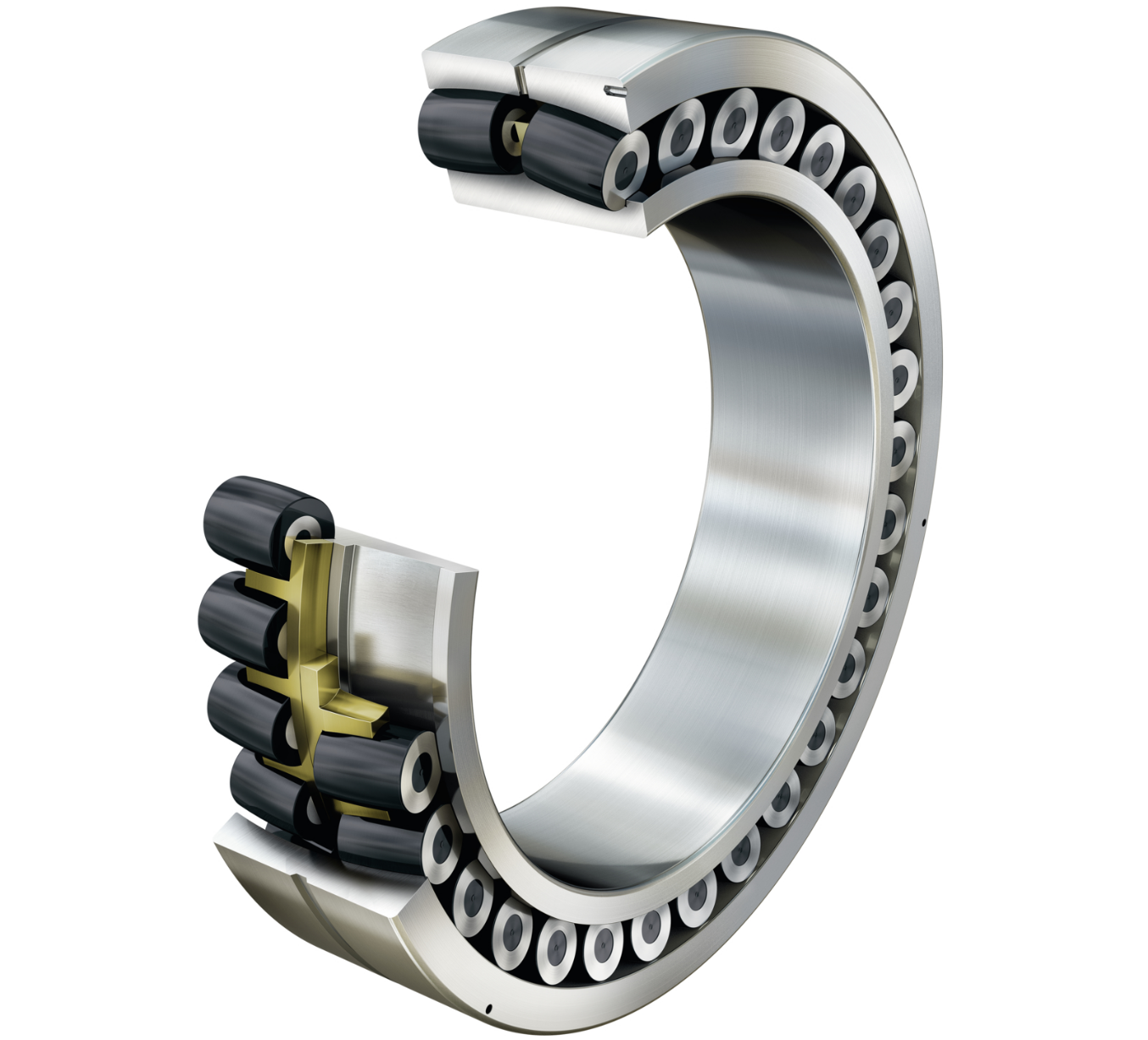}
  \includegraphics[width=0.28\linewidth]{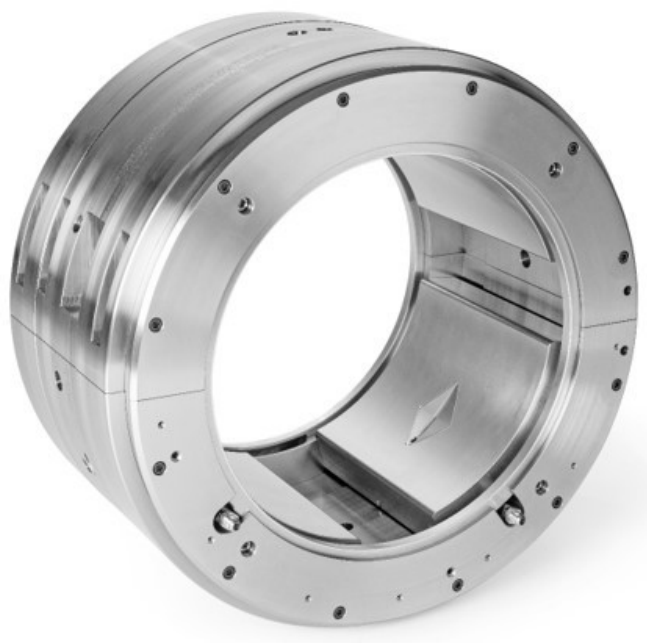}
 \caption{The left shows a Schaeffler roller bearing for the main shaft of a wind turbine \cite{schaeffler-bearing} while the right shows an example of a Miba tilting pad journal bearing used in turbines\cite{miba-pad-bearing}. In both examples sensors could be placed on the outside of the casing before putting the bearing in its mounting. 
 }
    \label{fig:raceway examples}
\end{figure}

Bearings are mounted in many different ways, one example is shown in \Cref{fig:raceway in mounting}. However, the raceway is usually a hollow cylinder and is fabricated as one solid piece, see \Cref{fig:raceway examples} for some examples. The raceway is then tightly fitted into the mounting, or if there is also an inner raceway then it is fitted over a shaft.

When the rollers press on the track, they emit elastic waves, which for high enough frequencies (> 10kHzs) are mostly trapped within the track \cite{randall2011rolling} due to the air gap that remains between the raceway and the mounting (or shaft).  
However, for convenience, sensors are usually placed on the bearing mounting, rather than the raceway itself, which does not always get a good signal for bearing defects, or other features. In many cases, the waves originally emitted into the raceway can take a long (transmission) path until reaching a sensor on the mounting. During this journey, the wave is highly distorted; this can make it difficult to recognize defects signals \cite{sawalhi2007enhancement}. There are methods which attempt to undo the effects of this path \cite{randall2021vibration} for impulsive signals, e.g. minimum entropy deconvolution \cite{sawalhi2007enhancement}. However, these can not be generalised to non impulsive signals; can enhance noise that is impulsive and have some difficulties in parameter choice such as window length \cite{antoni2006spectral}.  

What if we could mount sensors on, or near, a part of the raceway, as shown in \Cref{fig:raceway in mounting}? This has added complications as it needs to be planned into the design of the bearings, but, as show in this paper, it would also lead to many benefits such as a direct prediction of the stresses in bearings, and clear signals on defects, be them extended or localised. We can answer exactly what is possible to predict; what frequencies to use and where to place the sensors just considering that the raceway is approximately a hollow cylinder, as we do in the next section.

\subsection{Modal solution}\label{sec:general}

Below we show the simplest way to calculate elastic waves in the raceway, see \Cref{fig:bearing_geometry} for an illustration of the boundaries and domain where we calculate elastic waves.

Steel is well approximated as an isotropic material. Further, as even very high stresses of 1000 MPa only change elastic wave speeds by a few percent \cite{li2020ultrasonic}, the elastic waves within the raceway are well approximated by the linearised equations of elasticity in a homogeneous and isotropic solid \cite{marsden2012mathematical}. We also assume that stresses applied to the raceway boundaries are approximately axially symmetric, at least after averaging over some time period, which implies that the elastic waves are axially symmetric. 

The above allows us to write the small elastic displacement, for a harmonic angular frequency of  $\omega$ in terms of the Helmholtz potentials in the form  
\begin{equation}\label{eqn:helmdec}
    \boldsymbol u = \nabla \phi + \nabla \times (\psi \hat{\boldsymbol z}) 
\end{equation}
where $\phi$ and $\psi$ are the pressure and shear potentials respectively, and noting that the vector shear potential automatically satisfies the divergence free condition  $\nabla \cdot(\psi \hat{\boldsymbol z}) = 0$ when $\psi$ does not depend on $z$. The displacement in time can be easily calculated by taking an inverse Fourier transform, which is the same as integrating $\boldsymbol u  \ee^{- \ii \omega t}$ over $\omega$ in the convention used here.

The advantage of using the Helmholtz decomposition \eqref{eqn:helmdec} is that both potentials satisfy a Helmholtz equation:
\begin{equation}\label{eqn:helm-phi}
    \lap \phi + k_p^2 \phi = 0, \quad 
    \lap \psi + k_s^2 \psi = 0,
\end{equation}
where $k_p = \frac{\omega}{c_p}$ and $k_s = \frac{\omega}{c_s}$ are the wavenumbers of the P and S-waves, while $c_p$ and $c_s$ are the wave speeds which are related to the Lam\'e parameters $\lambda$ and $\mu$ through
\begin{equation} \label{eqn:Lame parameters}
    \rho c_p^2 = \lambda + 2\mu \quad \text{and} \quad 
    \rho c_s^2 = \mu,
\end{equation}
where $\rho$ is the mass density.



As we consider the raceway to be a thick-walled circular cylinder, we can reach simple solutions by using cylindrical coordinates $(r,\theta)$, which leads to solutions of \eqref{eqn:helm-phi} in the form
\begin{equation} 
\label{eqn:potentials}
\begin{aligned}
    & \phi(r,\theta) = \sum_{n=-\infty}^{\infty} \left(a_n \besselj_n(k_p r) + b_n \hankelh^{(1)}_n(k_p r)\right)\e^{\im n\theta },
\\
    & \psi(r,\theta) = \sum_{n=-\infty}^{\infty} \left(c_n \besselj_n(k_s r) + d_n \hankelh^{(1)}_n(k_s r)\right)\e^{\im n\theta },
\end{aligned}
\end{equation}
where $\besselj_n$ and $\hankelh^{(1)}_n$ are Bessel and Hankel functions of the first kind respectively. To deduce above you can use separation of variables \cite{martin2006multiple,propagation}, and the coefficients $a_n, b_n, c_n, d_n$ can be determined from boundary conditions.
Note that if the cylinder had no hole, then $b_n = d_n = 0$ and there would also be one less boundary to prescribe boundary conditions.

To prescribe boundary conditions we need the traction on the boundary in polar coordinates. In general the Cauchy stress tensor is given by
\begin{align}\label{eqn:constitutive}
    \vec \sigma = \lambda \mathrm{tr}(\vec \varepsilon) \mathbf{I} + 2\mu \vec \varepsilon
\end{align}
where $\boldsymbol \varepsilon = \frac{1}{2}\left(\nabla \vec u + \nabla \vec u^\mathrm{T}\right)$. The traction $\boldsymbol \tau$ on the outer boundary of a cylinder is given by 
\begin{equation}
    \boldsymbol \tau = \boldsymbol \sigma \cdot \hat{\boldsymbol r} = \sigma_{rr}\hat{\boldsymbol r} + \sigma_{r\theta} \hat{\boldsymbol \theta},
\end{equation}
where $\hat{\boldsymbol r}$ and $\hat{\boldsymbol \theta}$ are unit vectors along the directions that the radius and polar angle increase.  The traction on the inner boundary is given by $\boldsymbol \tau = - \boldsymbol \sigma \cdot \hat{\boldsymbol r}$ as the outward normal vector in this case is $-\hat{\boldsymbol r}$. See \cite{gould} for more details on stress tensors in polar coordinates.

Substituting \eqref{eqn:helmdec} into \eqref{eqn:constitutive} leads to  
\begin{align}\label{eqn:sig_rr}
    &\sigma_{rr} = \left(2c_s^2 k_p^2 - \omega^2\right)\rho \phi + 2\rho c_s^2 \left[\frac{\partial^2 \phi}{\partial r^2}+\frac{\partial}{\partial r}\left(\frac{1}{r}\frac{\partial \psi}{\partial \theta}\right)\right],
\\ \label{eqn:sig_rth}
    & \sigma_{r\theta} = -\rho \omega^2 \psi - 2\rho c_s^2\left[\frac{\partial^2 \psi}{\partial r^2}-\frac{\partial}{\partial r}\left(\frac{1}{r}\frac{\partial \phi}{\partial \theta}\right)\right].
\end{align}
See Appendix \ref{sec:app1} for derivation.

\section{Boundary conditions}\label{sec:bcs}

How much boundary data is needed to determine the potentials in \Cref{eqn:potentials}? To answer this question, consider the simpler case of how much data is needed to determine just the coefficients $a_n$ in a series $f(\theta) = \sum_n a_n \ee^{\ii n \theta};$ representing any square integrable periodic function $f(\theta)$ (almost everywhere). To determine $a_n$ we need to supply a function $f(\theta)$. Therefore, to determine $a_n$, $b_n$, $c_n$, and $d_n$ we must supply four functions. 

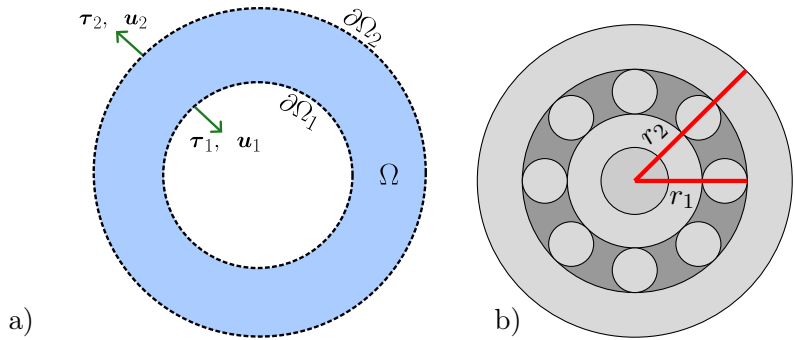
\begin{figure}[h]
\begin{center}
\includegraphics[width=0.3\linewidth]{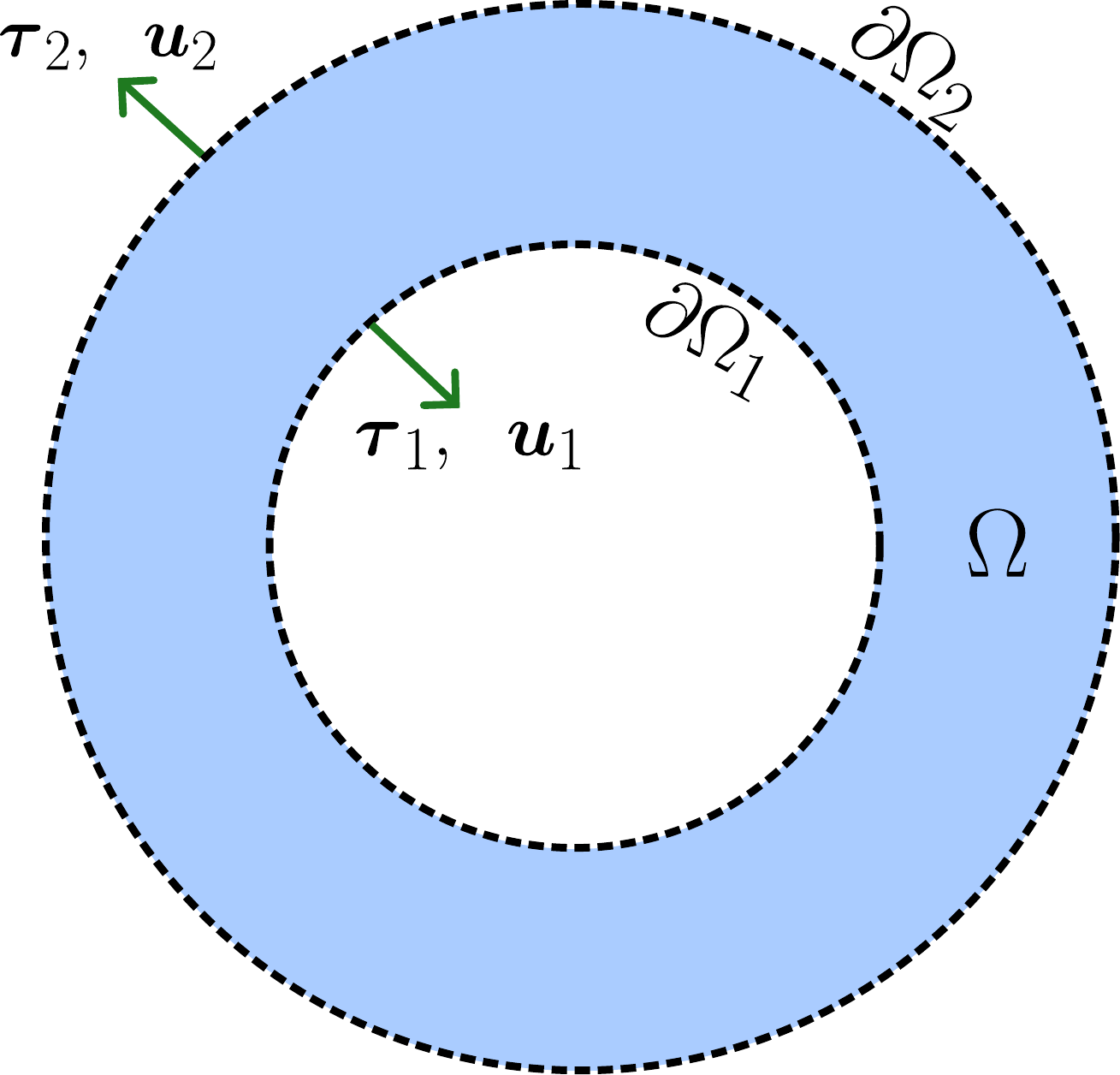} \hspace{0.4cm}
\begin{tikzpicture}[rotate=0,scale=0.74]
    \coordinate (O) at (0,0);
    \draw[fill=gray!30] (O) circle (2.8);
    \draw[fill=gray!80] (O) circle (2);
    \draw[fill=gray!30] (O) circle (1.2);
    \draw[fill=gray!40] (O) circle (0.6);
    \draw[fill=gray!30] (1.6,0) circle (0.4);
    \draw[fill=gray!30] (-1.6,0) circle (0.4);
    \draw[fill=gray!30] (0,-1.6) circle (0.4);
    \draw[fill=gray!30] (0,1.6) circle (0.4);
    \draw[fill=gray!30] (1.13137,1.13137) circle (0.4);
    \draw[fill=gray!30] (-1.13137,-1.13137) circle (0.4);
    \draw[fill=gray!30] (1.13137,-1.13137) circle (0.4);
    \draw[fill=gray!30] (-1.13137,1.13137) circle (0.4);
    \draw[ultra thick, red] (O) -- (2,0);
    \draw[ultra thick, red] (O) -- (1.9799,1.9799);
    \coordinate (A) at (0.85,0);
    \draw (A) node[below]{$r_1$};
    \coordinate (B) at (0.601,0.601);
    \draw (B) node[above, rotate=45]{$r_2$};
\end{tikzpicture}
\vspace{-0.6cm}

a) \hspace{5.8cm} b) \hspace{4.5cm}
\caption{Image a) on the left shows the general boundary conditions we consider. That is, we describe elastic waves in the domain $\Omega$, and will use some combination of the boundary data $\vec u_1, \, \vec u_2, \, \vec \tau_1, \, \vec \tau_2$ on the boundaries $\partial \Omega_1$ and $\partial \Omega_2$ which are defined by $r = r_1$ and $r = r_2$ respectively. The image b) on the right illustrates the geometry of one type of roller bearing, where the radii of the outer raceway are shown as an example. The methods of this paper could be used to predict the stresses between  either the rollers and outer raceway or the rollers and the inner raceway.}
\label{fig:bearing_geometry}
\end{center}
\end{figure}

For example, let us turn to \Cref{fig:bearing_geometry} and consider the outer raceway. To determine all the coefficients, it would be sufficient to have the boundary data of the displacement $\vec u$ and traction $\vec \tau$ on just one boundary, say at $r = r_1$, because
\[
\vec u(r_1,\theta) = u_r(r_1,\theta) \hat{\vec r} + u_\theta(r_1,\theta) \hat{\vec \theta},
\]
which is composed of two scalar functions in $\theta$, and likewise for $\vec \tau$. If the source of the elastic waves was also on the boundary $r = r_1$, then this would be an initial value problem, and we expect it then to be well-posed, and therefore this boundary data would completely determine the coefficients. 

Below we consider different combinations of boundary data and show how to determine the coefficients $a_n$, $b_n$, $c_n$, and $d_n$. In this work we do not discuss uniqueness of the solution, and simply verify that the solution is indeed well-posed where we expect it to be. That is, we expect the solution to become ill posed for low enough frequencies, and to be ill posed when approaching the diffraction limit, see \Cref{sec:app2} for details.

We call the \emph{forward problem} the case where the traction $\vec \tau$ on the inside and outside of the raceway are given. The name is just for convenience, and because knowing the traction on both boundaries often implies we know the source of the waves. However, this is still a boundary value, and could equally be considered to be an inverse problem. We consider this case first and then 
turn to more general boundary conditions, such as the case where the displacement $\vec u$ and traction $\vec \tau$ on the outer boundary are known, which we call the \emph{inverse problem}. Finally, we look at some example with stresses inspired by an operating bearing.

The inverse problem is of more practical importance, as it is possible to place sensors on the outside of the raceway for the case shown in \Cref{fig:bearing_geometry}. In \Cref{sec:measured points} we discuss the details on taking boundary data from a, possibly small, finite number of measurements at specific points on the boundary.

\subsection{Traction boundary conditions: the forward problem}\label{sec:trac}
Here we consider prescribing only traction boundary conditions on both the boundaries of the raceway, see \Cref{fig:bearing_geometry} for an illustration. 


Let the traction on the boundary $r=r_1$ be $\boldsymbol \tau^1(\theta)$, and the traction on the boundary $r = r_2$ be $\boldsymbol \tau^2(\theta)$, be given by;
\begin{align}\label{eqn:trac_inner}
    & \vec \tau^1(\theta) = 
    -p^1(\theta)\hat{\boldsymbol r} - s^1(\theta) \hat{\boldsymbol \theta},
    \quad \vec \tau^2(\theta)  
    =  p^2(\theta)\hat{\boldsymbol r} + s^2(\theta) \hat{\boldsymbol \theta},
\end{align}
where the negative sign in the top equation is due to the unit normal being $-\hat{\boldsymbol r}$. 

To solve the problem now we need the Fourier series representation:
\begin{align}
    & p^j(\theta) = \sum_{n=-\infty}^{\infty} p^j_n \e^{\im n\theta}, \quad 
    s^j(\theta) = \sum_{n=-\infty}^{\infty} s^j_n \e^{\im n\theta}, \quad \text{for} \;\; j=1,2,
\end{align}
then substitute the potentials \eqref{eqn:potentials} into \eqref{eqn:sig_rr} and \eqref{eqn:sig_rth} and then substitute the result into \eqref{eqn:trac_inner}. Using that the $\ee^{\ii n \theta}$ in the Fourier series are orthonormal leads to the matrix equation
\begin{equation}\label{eqn:m_for}
    \mathbf{M}^{\mathrm{for}}_n \boldsymbol a_n = \boldsymbol f_n,
\end{equation}
for the mode $n$; where, 
\[
\boldsymbol a_n = [a_n,b_n,c_n,d_n]^{\mathrm{T}} \quad \text{and}\quad
\boldsymbol f_n = [p_n^1,s_n^1,p_n^2,s_n^2]^{\mathrm{T}}.
\]
The expressions for the components of the 4$\times$4 matrix $\mathbf{M}^{\mathrm{for}}_n$ involve known special functions and can be found in \Cref{sec:app3}. If the matrix is well conditioned, then we can solve \eqref{eqn:m_for} for the coefficients $\boldsymbol a_n$. In practice, we numerically check whether $\mathbf{M}^{\mathrm{for}}_n$ is well conditioned for $n=0$, and then increase $|n| = 1, 2, \ldots$ until $\mathbf{M}^{\mathrm{for}}_n$ is not well conditioned. Further details on when this problem is ill-posed can be found in \Cref{sec:what is measureable} and \Cref{sec:app2}.

\subsection{Data on only one boundary: the inverse problem}\label{sec:mixed}

In practice it is not possible to know the traction on both boundaries. For example in the image on the right of \Cref{fig:bearing_geometry} it is clearly not feasible to have sensors on the boundary $r= r_1$, however the boundary $r=r_2$ is often approximately traction free, due to the small air gap between the raceway and mounting. If we place ultrasonic sensors on the boundary $r = r_2$ then we would also know the displacement $\vec u_2$. See \Cref{fig:bearing_measure} for an illustration. In the next section we discuss how to deal with a finite number of sensors, but for the remainder of this section we consider that both the traction and displacement are known on the boundary $r=r_2$.

Analogous to the previous section, we write the displacement on the outer-boundary as a Fourier series
\begin{align*}
    \boldsymbol u(r_2,\theta) = \vec u^2(\theta) = \sum_n u^{(r)}_n \hat{\boldsymbol r} \e^{\im n \theta} +  \sum_n u^{(\theta)}_n \hat{\boldsymbol \theta} \e^{\im n \theta},
\end{align*}
then by substituting the potentials \eqref{eqn:sig_rth} \eqref{eqn:sig_rr} into the expression for $\boldsymbol u$  in \eqref{eqn:helmdec}, and then substituting the result into the above leads to two separate equations. Combining these two equations with the two equations for the traction boundary data $\vec \tau^2$ from the previous section, and again using that the modes of the Fourier series are orthogonal leads to another 4$\times$4 matrix equation:
\begin{align}\label{eqn:m_inv}
    \mathbf{M}^{\mathrm{inv}}_n \vec a_n = \vec u_n.
\end{align}
If $\mathbf{M}^{\mathrm{inv}}_n$ is well conditioned, then we can solve for $\vec a_n$, and this solution will also solve the forward problem \eqref{eqn:m_for}. An example of solving a transient point force is given in \Cref{sec:example transient}






\subsection{Measured points on the boundary}
\label{sec:measured points}

Let us start by summarising our results so far. In \Cref{sec:bcs} we showed how to form a system
\begin{equation} \label{eqn:modal-system}
    \mathbf{M}_n \vec a_n = \vec f_n,
\end{equation}
for some given choice of boundary conditions, where the vector $\vec f_n$ contains the Fourier modes from measurements on the boundaries, i.e. the measured displacement and/or traction, and $\mathbf{M}_n$ is a known 4-by-4 matrix that depends on the type of boundary conditions. In practice, we do not have direct access to the Fourier modes of the boundary data $\vec f_n$, but instead measure the elastic wave displacement at specific points on the boundary. That is, by summing up the Fourier modes on both sides of the modal system \eqref{eqn:modal-system} we obtain: 
\begin{equation} \label{eqn:boundary system continuous}
    \sum_n \ee^{\ii n \theta} \mathbf M_n \vec a_n =  \sum_n \ee^{\ii n \theta} \vec f_n,
\end{equation}
where $\vec y(\theta) := \sum_n \ee^{\ii n \theta} \vec f_n$ represents all the boundary data as a function of $\theta$. In practice we may measure $\vec y(\theta)$ at specific angles $\theta$ and from this want to obtain the $\vec a_n$. 

A sophisticated approach would consider that $\vec y(\theta)$ is some statistical distribution that is estimated from measured data. However, in this work we want to keep the presentation as simple as possible. So instead, we show how to rewrite the system \eqref{eqn:boundary system continuous} in terms of a finite number of deterministic measurements on the boundaries.

 First note that $\vec y(\theta)$ covers two different boundaries, each of which could be sampled at different angles $\theta$. To accommodate this we rewrite \eqref{eqn:boundary system continuous} sampled at discrete angles
\begin{equation} \label{eqn:boundary system discrete}
    \sum_n\begin{bmatrix}
    \ee^{\ii n \theta_{m_1}^1} \mathbf M_n^1
    \\
    \ee^{\ii n \theta_{m_2}^2} \mathbf M_n^2
    \end{bmatrix}
     \vec a_n = 
     \begin{bmatrix}
         \vec y_\inv^1(\theta_{m_1}^1)
         \\
         \vec y_\inv^2(\theta_{m_2}^2)
     \end{bmatrix},
\end{equation}
where we evaluate $m_1 = 1,2,\ldots, M_1$ to iterate over the $M_1$ measured points $\theta_1^1, \theta_1^2, \ldots, \theta_1^{M_1}$. Likewise, we evaluate $m_2 = 1,2,\ldots, M_2$. However, to facilitate implementation, we want to iterate over just one index $m$, rather than $m_1$ and $m_2$, which leads us to rewrite the left of \eqref{eqn:boundary system discrete} in the form 
\begin{align} \label{eqn:boundary system discrete 2}
     & \sum_n \mathbf E_{m n} \vec a_n =  
     \begin{bmatrix}
         \chi_{m}^1 \vec y_\inv^1(\theta_{m}^1)
         \\
         \chi_{m}^2 \vec y_\inv^2(\theta_{m}^2)
     \end{bmatrix} \quad \text{with} \quad
      \mathbf E_{m n} := 
     \begin{bmatrix}
       \chi_{m}^1 \ee^{\ii n \theta_{m}^1} \mathbf M_n^1
        \\
       \chi_{m}^2 \ee^{\ii n \theta_{m}^2} \mathbf M_n^2
    \end{bmatrix}, 
\end{align}
where $\chi_{m}^1 = 1$ if $1 \leq m \leq M_1$, and otherwise  $\chi_{m}^1 = 0$, and $\chi_{m}^2$ has the analogous definition. We note the technicality that $\theta_m^j$ is not defined if $\chi_m^j = 0$, which we can remedy by setting $\theta_m^j = 0$ when $\chi_m^j = 0$.


To solve \eqref{eqn:boundary system discrete 2} it is best to rewrite it in the block matrix form: 
\begin{equation}\label{eqn:bounday block discrete}
 \mathbf E \vec a = \vec y,
\end{equation}
where $\mathbf E$ is a block matrix with the matrix block components $\mathbf E_{mn}$,  $\vec a$ is a block vector  from vertically stacking the vectors $\vec a_n$, and likewise $\vec y$ is a block vector which results from vertically stacking the vectors on the right side of \eqref{eqn:boundary system discrete 2}. 

Finally, for \eqref{eqn:bounday block discrete} to have a unique solution for $\vec a$ then the number of modes $N$ considered for $\vec a_n$ has to satisfy $N \leq M_1$ and $N \leq M_2$.



\section{Priors and recovering the load from the rollers}
\label{sec:prior method}

The methods shown in \Cref{sec:bcs} make no assumptions about the boundary conditions. If we make no assumptions about the internal geometry, or sources of the elastic waves, then we may need a lot of sensors to obtained detailed prediction of the boundary data, as shown in \Cref{sec:ex-localised force}. To use a small number of sensors we need to provide some information, which we call priors. 

For example, using a tachometer (a revolution counter), together with the design specifications of the roller bearing, we would know approximately the speed of the rollers and their contact points. We show later that this in itself is a powerful prior. 

\subsection{Linear priors} \label{sec:linear priors}
Any prior information about the source of waves, such as a known number of roller bearings, will allow us to parameterise $\boldsymbol a$ in some way. For instance, a linear parameterisation:
\begin{equation} \label{eqn:block prior}
    \vec a = \mathbf{B} \vec x + \vec{c},
\end{equation} 
where the matrix $\mathbf{B}$, and bias vector $\vec c$, are known from the prior information, while $\vec x$ is now the unknown. We assume that $\mathbf B$ is full-column rank. For the above to be a restriction on $\vec a$ the matrix $\mathbf B$ has to have more rows than columns. We will show later how a linear relationship between $\vec a$ and $\vec x$ covers many important cases.


Substituting \eqref{eqn:block prior} into the block modal equation \eqref{eqn:bounday block discrete} then leads to 
\begin{equation}
    \mathbf{E}\mathbf{B}  \boldsymbol x + \mathbf{E}\boldsymbol{c}= \vec y,
\end{equation}
where we use a pseudo inverse to obtain a solution
 \begin{equation} \label{eqn:pseudo inverse prior}
    \vec x_\star = ( \mathbf{E}\mathbf{B})^+(\vec y - \mathbf{E}\textbf{c}).
\end{equation}
For the above to give a unique solution, we have also assumed that $\mathbf E$ is full rank. Substituting the above into \eqref{eqn:block prior} leads to
\begin{equation} \label{eqn:block prior result}
    \vec a_\star = \mathbf{B} ( \mathbf{E}\mathbf{B})^+ (\vec y - \mathbf{E} \textbf{c})+ \vec{c}.
\end{equation}
The solution $\vec a_\star$ would be equal to $\vec a$ when $\mathbf E \mathbf B$ is a square matrix, otherwise $\vec a_\star$ is a least squares approximation to $\vec a$. We give an example for roller bearings in \Cref{sec:example-loading-profile}.

Note that while the inverse of $\mathbf{E}$ could be ill defined, the pseudo inverse of $\mathbf{E} \mathbf{B}$ could be well defined, as we expect the dimension of $\vec x$ to be much smaller than the dimension of $\vec a$. 

\subsection{Prior due to boundary conditions}
\label{sec:priors due to boundary conditions}

One of the most general ways to have a linear prior, as shown in \Cref{sec:linear priors}, is to have a linear basis for the boundary conditions. For example, the boundary where the rollers make contact with the raceway leads to a basis as shown in \Cref{sec:roller bearings}.

In this section we use $\Mfor_n$ and $\Efor$ to represent the modal matrix and block matrix in \eqref{eqn:modal-system} and \eqref{eqn:bounday block discrete} for the boundary conditions for which we have prior knowledge. The "$\for$" in $\Mfor$ stands for forward problem\footnote{Although it is debatable what is a forward or inverse problem here.}. 
 We use $\Minv_n$ and $\Einv$ for problems which are ill posed, where "$\inv$" represents inverse problem.



To completely determine the elastic waves within the bearing would require the boundary data $\vec y^1_\for(\theta)$ and $\vec y^2_\for(\theta)$, each of which can be written in terms of scalar functions in the form 
\begin{equation}
    \vec{y}^1_\for(\theta) = 
    \begin{bmatrix}
        p^1(\theta) \\
        s^1(\theta)
    \end{bmatrix}
    \quad \text{and} \quad 
    \vec{y}^2_\for(\theta) = 
    \begin{bmatrix}
        p^2(\theta) \\
        s^2(\theta)
    \end{bmatrix}.
\end{equation}
For example, if the boundary data $\vec{y}^1_\for(\theta)$ represents the traction (see \Cref{fig:bearing_geometry}), then $p^1(\theta)$ and $s^1(\theta)$ would represent the pressure and shear force as a function of the angle $\theta$.

To reach a linear prior \eqref{eqn:block prior}, we assume there is a known basis for the boundary data:
\begin{align} \label{eqn:boundary basis}
    & \vec y^1_\for(\theta) = \sum_{\ell=0}^{L_1} x^1_\ell \vec y^1_\ell(\theta) + \vec b^1(\theta), \quad 
    \vec y^2_\for(\theta) = \sum_{\ell = 0}^{L_2} x^2_\ell \vec y^2_\ell(\theta) + \vec b^2(\theta),
\end{align}
where the $\vec y^j_\ell(\theta)$ and $\vec b^j(\theta)$ are known, and the $x^j_\ell$ are, for now, unknown.
Each of these functions can be decomposed in  Fourier modes:
\begin{align}
    \vec y_\ell^j(\theta) = \sum_n \vec f_{\ell n}^j \e^{\im n \theta}
    \quad \text{and} \quad
    \vec b^j(\theta) = \sum_n \vec b_{n}^j \e ^{\im n \theta}, \quad \text{for} \;\; j=1,2.
\end{align}

Using the above, we can write the boundary conditions for one mode in the form:
\begin{equation} \label{eqn:modal prior}
    \Mfor_n \boldsymbol a_n = 
    \begin{bmatrix}
        \mathbf{F}_n^1
        \boldsymbol x^1 \\
        \mathbf{F}_n^2
        \boldsymbol x^2
    \end{bmatrix}
    +
    \begin{bmatrix}
        \vec b_n^1
         \\
        \vec b_n^2
    \end{bmatrix}, 
\end{equation}
where we define
\begin{align}
    & \mathbf{F}_n^j= \begin{bmatrix}
        \vec{f}^j_{1n} & \vec{f}^j_{2n} & \cdots & \vec{f}^j_{L_j n}
        \end{bmatrix},
\end{align}
so that $\mathbf{F}_n^j \vec x^j = \sum_\ell \vec f_{\ell n}^j x_\ell^j$. To write the above in a block matrix form we define
\begin{align} \label{def:block priors}
        \vec b_n = \begin{bmatrix}
            \vec{b}^1_n \\
            \vec{b}^2_n
        \end{bmatrix}, \quad 
        & \vec x = \begin{bmatrix}
            \vec{x}^1 \\
            \vec{x}^2
        \end{bmatrix},          
        \quad \text{and} \quad 
    \mathbf{F}_n = 
    \begin{bmatrix}
        \mathbf{F}_n^1 & 0 \\
        0   & \mathbf F_n^2
    \end{bmatrix}, 
    \quad \text{so that} \quad \mathbf F_n \vec x = \begin{bmatrix}
    \mathbf{F}_n^1
    \boldsymbol x^1 \\
    \mathbf{F}_n^2
    \boldsymbol x^2
    \end{bmatrix},
\end{align}
and then rewrite \eqref{eqn:modal prior} in a block form
to obtain
\begin{align} \label{eqn:block prior boundary}
   & \Mfor \vec a = \mathbf{F}^\for \vec x + \vec b^\for, \quad  \text{with} \quad
     \mathbf{F}^\for = \begin{bmatrix}
        \vdots\\
        \mathbf{F}_{-1}\\
        \mathbf{F}_0\\
        \mathbf{F}_1\\
        \vdots
    \end{bmatrix}
    \quad \text{and} \quad 
    \vec b^\for = \begin{bmatrix}
        \vdots\\
        \vec{b}_{-1}\\
        \vec{b}_0\\
        \vec{b}_1\\
        \vdots
    \end{bmatrix},
\end{align}
where $\Mfor$ is a block diagonal matrix with $\Mfor_n$ on the diagonals, and we have added the superscript ``$\for$'' to emphasize that these above quantities are related to the forward problem. Note that if we knew the boundary data of the forward problem we would have $\vec f^\for = \mathbf F^\for \vec x + \vec b^\for$.

Finally, we take the inverse of $\Mfor$ on both sides of \eqref{eqn:block prior boundary} to obtain
\begin{equation} \label{eqn:block prior boundary result}
     \vec a = (\Mfor)^{-1}\mathbf{F}^\for \vec x + (\Mfor)^{-1} \vec b^\for,
\end{equation}
where we have assumed that the type of boundary conditions that lead to $\Mfor$ lead to a well conditioned problem so that calculating the inverse $(\Mfor)^{-1}$ is stable and well defined.


The restriction \eqref{eqn:block prior boundary result} on $\vec a$ now matches the abstract form given by \eqref{eqn:block prior result M}, where by comparison we obtain:
\begin{equation} \label{eqn:B and c boundary data}
    \mathbf{B}^\for = (\mathbf{M}^{\text{for}})^{-1} \mathbf{F}^\for
\quad\text{and} \quad
\vec{c}^\for = (\mathbf{M}^{\text{for}})^{-1} \vec b^\for.
\end{equation}
We can use the above restriction to solve for $\vec a$ even when given incomplete boundary data.
Let us write this out in full for clarity.

Let $\mathbf E = \Einv$ and $\vec y = \vec y^\inv$ in \eqref{eqn:bounday block discrete} to indicate that calculating $(\Einv)^{-1}$ is either ill-posed or that the measured $\vec y^\inv$ is incomplete boundary data. Our aim is now to solve
\begin{equation} \label{eqn:block modal inverse}
    \Einv \vec a = \vec y^\inv.
\end{equation}
Using the result \eqref{eqn:block prior result} together with the substitutions \eqref{eqn:B and c boundary data}  leads to the solution
\begin{equation} \label{eqn:block prior result M}
    \vec a_\star = \mathbf{B}^\for ( \Einv \mathbf{B}^\for)^+ (\vec y^\inv - \Einv \textbf{c}^\for)+ \vec{c}^\for.
\end{equation}

It is likely easier to understand this result, and its consequences, with a concrete example which we provide for roller bearings in \Cref{sec:roller bearings}. Nonetheless, let us consider some here important features of this solution. 

To simplify the discussion here, let us assume that the number of boundary measurements $\vec y^\inv$ is equal to the number of unknowns in $\vec x$ so that $\vec a_\star = \vec a$ and $\Einv \mathbf{B}^\for$ is a square matrix. So if we have a representation for $\vec a$ that uses a small number of basis elements $L_1$ and $L_2$ in \eqref{eqn:boundary basis}, then we need only a small number of measurements in $\vec y^\inv$ to obtain the unique solution $\vec a$. In \Cref{sec:roller bearings} we show how assuming a smooth loading of a roller bearing leads to small values for $L_1$ and $L_2$. To further emphasize this point,  note that the resolution of the solution is governed by the number of modes $N$ in $\vec a$. That is, the block vector $\vec a$ is formed of the vectors $\vec a_n$ with $n$ having $N$ possible values. For a fixed number of basis' $L_1$ and $L_2$ in \eqref{eqn:boundary basis} we can increase $N$ and still obtain the unique solution $\vec a$ as long as $\Mfor$ in \eqref{eqn:block prior boundary result} continues to be well conditioned. This is why the images in \Cref{fig:true and prior stress} have such high resolution, despite having only 3 sensors.

\subsection{One traction free boundary}
\label{sec:example traction free boundary}
Here we give an example which is typical for bearings: the traction on one boundary is known, and the boundary in contact with the roller bearings has a basis function. For example the outside boundary in Figure \ref{fig:bearing_geometry} could be traction free. 

For this case, we need to make a small adjustment to the prior method shown in the previous section. Here the boundary data is of the form 
\begin{align}
    & \vec y^1_\for(\theta) = \sum_{\ell=0}^{L_1} x^1_\ell \vec y^1_\ell(\theta), \quad 
    \vec y^2_\for(\theta) = \vec b^2,
\end{align}
so that $\vec b^1(\theta) = \vec 0$ and $\vec x^2 = \vec 0$.

Following the same steps shown in the previous section, we would need to make a small adjustment by redefining 
\begin{equation}
    \vec x = \vec{x}^1, 
        \quad \text{and} \quad 
    \mathbf{F}_n= \begin{bmatrix}
     \mathbf{F}_n^1 \\
        0 
    \end{bmatrix}.
\end{equation}

\section{Roller bearings and the loading profile} \label{sec:roller bearings}

Here we develop an application for roller bearings that shows the great potential of describing the elastic waves in the raceway in more detail.

\begin{figure}[h!] 
\centering 
    \begin{minipage}[b]{.45\linewidth}
        \centering\includegraphics[width=0.92\linewidth]{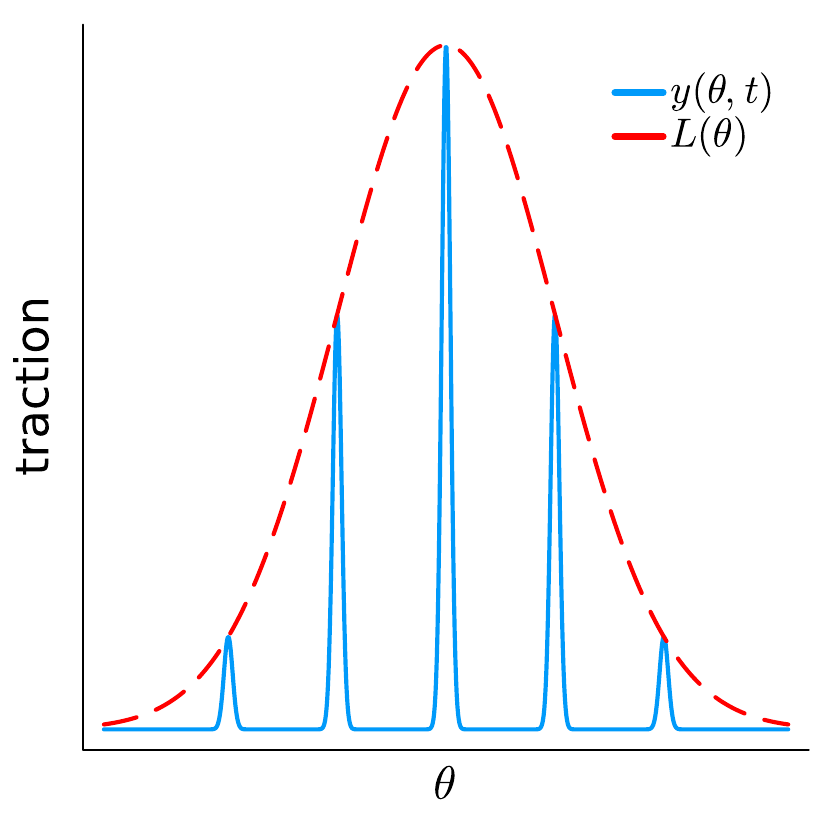}
        %
    \end{minipage} \hspace{0.2cm}%
    \begin{minipage}[b]{.44\linewidth}
        \centering
        \definecolor{lightblue}{rgb}{0.19, 0.55, 0.91}
        \begin{tikzpicture}
            \clip (-3,-3.5) rectangle (3,1);
            \filldraw[cyan!10] (-3,-3.5) rectangle (3,1);
            \coordinate (O) at (0,4);
            \draw[fill=gray!30] (O) circle (6.8);
            \draw[fill=gray!80] (O) circle (5);
            \draw[fill=gray!30] (O) circle (3.7);
            \draw[fill=gray!30] (0,-0.35) circle (0.65);
            \draw[fill=gray!30, shift={(1.7,0)}, rotate=-68] (0,-0.35) circle (0.65);
            \draw[fill=gray!30, shift={(-1.7,0)}, rotate=68] (0,-0.35) circle (0.65);
            \filldraw[red,shift={(0,4)}, rotate=240, opacity=0.2] (0,0) - - (6.8,0) arc (0:60:6.8) - - cycle;
            \draw[fill=gray!40] (O) circle (1.9);
            \draw[ultra thick, lightblue,->] (0,-0.35) -- (0,-2.5);
            \draw[ultra thick, lightblue,->] (-1.3754,-0.1311) -- (-1.8,-1.16555);
            \draw[ultra thick, lightblue,->] (1.3754,-0.1311) -- (1.8,-1.16555);
        \end{tikzpicture} \vspace{1cm}
        
    \end{minipage}%

    \vspace{-0.5cm}
    \caption{The graph on the left shows the stress on the raceway boundary $y(\theta,t)$. Each blue spike is the result of one roller being in contact with the raceway. As the rollers move in time, the blue spikes in this graph also move, but they all trace the same curve $L(\theta)$, where we assume the load supported by the bearing does not change in time. The image on the right illustrates how load is transmitted through the rollers.}
    \label{fig:loading}
\end{figure}
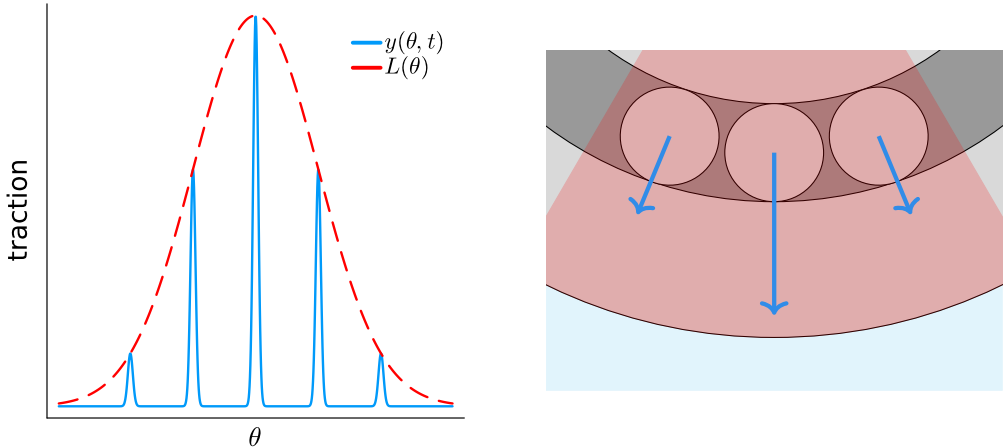

Consider a roller bearing as shown on the left of \Cref{fig:raceway examples} and on the right of \Cref{fig:bearing_geometry}. Any load applied to the shaft in the middle of the bearing, or applied to the outer raceway, will be transmitted through the rollers themselves with each roller in contact with only a small region of the raceway, as illustrated in \Cref{fig:loading}. We do not need to know the exact shape of this small contact region if the goal is just to measure the overall load passed through the bearing; as long as the contact region is small compared to the bearing geometry. Below we show how knowing the rotation speed of the bearing, number of bearings, and their contact points, can lead us to predict the load transmitted through each bearing with very few sensors. 

We make a number of simplifying assumptions, which can improved on in future work. First, in practice, rollers slip as they go around \cite{randall2021vibration}, making their contact points better described as a random variable. Further, the contact points of the bearings with the raceway are more accurately modelled as Hertzian contacts \cite{contactmechanics}. However, here we show only how to use deterministic priors both for simplicity but because it is necessary to develop the deterministic version first before developing more precise models with random variables. Second, we assume the bearing is rotating at a constant speed $\Omega$. The framework we present can accommodate any change in rotation speed, but the conclusions shown below would need to be adjusted.

\subsection{Static Loading profile} 
\label{sec:static-loading-profile}

Imagine the bearings are loaded just due to gravity, or some other static forces on the shaft, mounting, or casing.
We use the function $L(\theta)$ to denote the stress transmitted through a roller when it is in contact with the raceway at an angle $\theta$. We call $L(\theta)$ the loading profile and in this section assume it is independent of time. So this excludes enviroment effects, for example.

Assume that $L(\theta)$ represents the radial stress for simplicity, and that the contact region of the roller is small, then we can write that the radial stress on the boundary $y(\theta,t)$ of the raceway is given by 
\begin{equation} \label{eqn:bearing point forcing static}
    y(\theta,t) = L(\theta) d(\theta,t), \quad \text{where} 
    \quad d(\theta,t) = \sum_{s = -\infty}^\infty\delta \left ((\theta - \Omega t)Z + 2\pi s \right),
\end{equation}
where $Z$ is the number of roller bearings, $\Omega$ their angular speed, $t$ is time, and the function $\delta(\theta)$ represents the stress distribution due to one roller. See \Cref{fig:loading} for an illustration. A form similar to \eqref{eqn:bearing point forcing static} for the stress on the boundary was introduced in \cite{mcfadden1984model} and \cite{mcfadden1985vibration}.


The function $d(\theta,t)$ moves the contact points of the bearing as time passes, and assumes that the $Z$ bearings have the same distance between each other. We assume that each $\delta \left ((\theta - \Omega t)Z + 2\pi s \right)$ when integrated over $\theta$ is equal to 1, so that the magnitude of the load transmitted is always $L(\theta)$, no matter the shape of the function $\delta(\theta)$. We could, for example have the function $\delta$ be as scaled Dirac delta function. However to avoid Gibbs phenomena it is best to use a Gaussian function:
\begin{equation}
  \delta \left (x\right) =   \frac{Z}{\sigma} \ee^{ - \pi  x^2/\sigma^2},
\end{equation}
where $\sigma$ is the standard deviation of the contact spread. In the limit of $\sigma \to 0$ the above $\delta$ would become a scaled Dirac delta. Naturally, we could use other type of contact points, but when the aim is to measure $L(\theta)$ we do not need to model precisely the contact region. 

The function $d(\theta,t)$ is periodic in time with period $T = 2\pi / (Z \Omega)$, which means we can write $d(\theta,t)$ in terms of its Fourier series in time, which (after some calculations) is given by 
\begin{equation} \label{eqn:contact points fourier}
d(\theta,t) = \frac{Z}{2 \pi} \sum_{m = -\infty}^\infty  \ee^{-\pi \sigma^2 m^2} \cos(m Z [\theta - \Omega t] ), \quad \text{where} \quad \omega_m = m Z \Omega.
\end{equation}

The loading profile $L(\theta)$ is also $2\pi$ periodic in $\theta$, so we use a Fourier series representation: 
\begin{equation} \label{eqn:loading profile fourier series}
   L(\theta ) = \sum_n c_n \ee^{\im n  \theta},
\end{equation}
which substituted into \eqref{eqn:bearing point forcing  static} together with \eqref{eqn:contact points fourier}, and after some calculations, leads to 
\begin{equation} \label{eqn:roller loading to coefficients}
    y(\theta,t) = \sum_{n,m} f_n(\omega_m)\ee^{\im n \theta} \ee^{ - \im \omega_m t}, 
    \quad \text{with} \quad f_n(\omega_m) = \frac{Z}{2 \pi}   c_{n- m Z} \ee^{-\pi \sigma^2 m^2},
\end{equation}
which matches the notation from the previous sections. 

Using the above, together with the prior method shown in \Cref{sec:priors due to boundary conditions} and \Cref{sec:example traction free boundary}, we can recover the coefficients of the loading profile $c_n$ by measuring the displacement on the boundary of the raceway that is traction free. To do so, we first identify the unknowns $x_\ell = c _\ell$, then the matrix $\mathbf F_n^1$ would be full of zero except for the column number $\ell = n - m Z$ which would be 
\[
\vec f^1_{\ell, n} = \frac{Z}{2\pi}\begin{bmatrix}
    \ee^{-\pi \sigma_P^2 m^2} \\
    \mu_S \ee^{-\pi \sigma_S^2 m^2}
\end{bmatrix},
\]
where we assume the contact force distribution for the pressure $\sigma_P$ is potentially different than the contact force for the shear $\sigma_S$, and we also assume that if the pressure is known, then the shear is known as consequence.  

In the examples section, we show that four different sensors are needed to recover the load accurately, in steel, if the loading profile is smooth, as shown in \Cref{sec:example-loading-profile}. This is because a smooth loading profile implies that the series \eqref{eqn:loading profile fourier series} needs few terms to converge.

\subsection{Quasi-Static Loading profile}
The most general force due to the rollers bearings on the raceway is given by 
\begin{equation} \label{eqn:bearing point forcing}
    y(\theta,t) = L(\theta,t) d(\theta,t), 
\end{equation}
instead of \eqref{eqn:bearing point forcing}. By taking the Fourier transform of both sides and using \eqref{eqn:contact points fourier} with convolution theorem we obtain 
\begin{align} \label{eqn:bearing point forcing fourier}
    \hat y(\theta,\omega) 
    &= \frac{Z}{2\pi} \sum_m  \ee^{-\pi \sigma^2 m^2}\left(\hat L(\theta, \omega - \omega_m)  \ee^{\im m Z \theta} + \hat L(\theta, \omega + \omega_m)  \ee^{-\im m Z \theta}\right), 
\end{align}
where $\hat y(\theta,\omega)$ and $\hat L(\theta, \omega)$ are the Fourier transforms of $y(\theta,t)$  and $L(\theta, t)$ respectively. 


The form \eqref{eqn:bearing point forcing fourier} would not be a useful prior if we knew nothing about $\hat L$. However, there is a useful and practical assumption that the loading is quasi-static, i.e. does not change rapidly. The simplest scenario being that $\hat L(\theta, \omega) \approx 0$ for $|\omega| > Z \Omega$, in which case the sum in \eqref{eqn:bearing point forcing fourier} reduces to just one value for $m$, leading to
\begin{equation} \label{eqn:bearing point forcing quasi-static}
    \hat y(\theta,\omega) = 
    \frac{Z}{2\pi} \ee^{-\pi \sigma^2 m^2} \left(\hat L(\theta, \omega - \omega_m)  \ee^{\im m Z \theta} + \hat L(\theta, \omega + \omega_m)  \ee^{-\im m Z \theta}\right), 
\end{equation}
where $m = \Big\lfloor \frac{\omega}{Z \Omega} \Big \rceil$, with $\lfloor x \rceil$ being equal to $x$ rounded to the nearest integer. 

Analogous to the previous section, we decompose $\hat y$ and $\hat L$ in their Fourier modes to obtain
\begin{equation}\label{eqn:boundary to loading modes quasi static}
f_n(\omega) = \frac{Z}{2\pi} \ee^{-\pi \sigma^2 m^2} c_{n - m Z}(\omega-\omega_m).
\end{equation}

\section{What is measurable}
\label{sec:what is measureable}

\begin{figure}[h]
    \centering
    \includegraphics[width=0.49\linewidth]{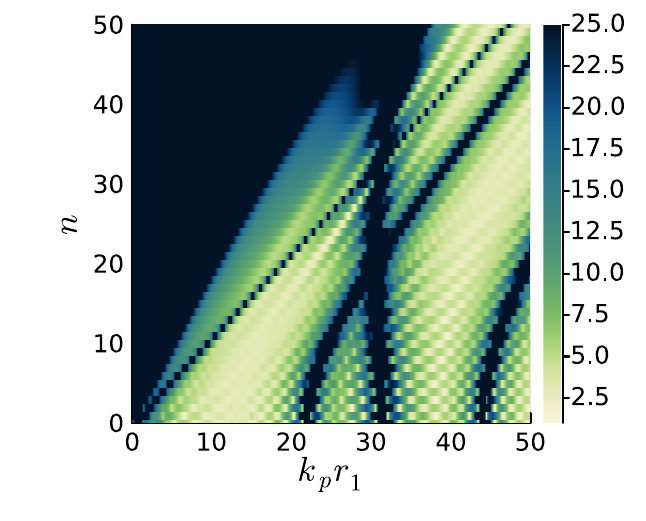}
    \includegraphics[width=0.49\linewidth]{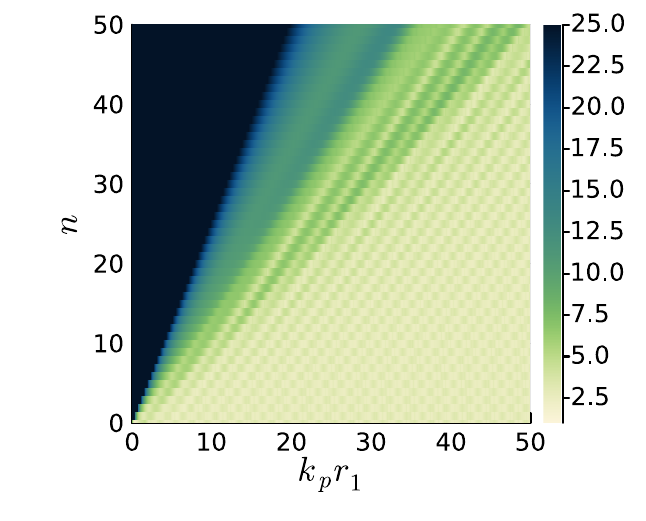}
    \vspace{-0.5cm}
    
    \caption{The above heatmaps show the condition number cond$\,\mathbf{M}_n^\for$ on the left and cond$\,\mathbf{M}_n^{\text{inv}}$ on the right, after non-dimensionalisation. In the dark regions the errors in the boundary data can be amplified by 20 times. Note that for any fixed $k_p r_1$ if we keep increasing $n$ the system will at some point become ill-conditioned. The dark cross and lines on the forward problem are resonant modes.
    The parameters used are given in \Cref{tab:params}.
}
    \label{fig:condition-heatmaps}
\end{figure}

It is not always possible to robustly estimate the stresses between the rollers, or other elements with elastic waves. There are two main phenomena that cause this: 1) resonance and 2) the diffraction limit. Numerically, we observe a relationship between the stability of the inverse problem and the frequency $\omega$. In particular, we find that the numerical stability of the inverse problem increases as we increase the frequency, see \Cref{fig:condition-heatmaps}. This relationship between numerical stability and frequency is a well-established phenomenon in inverse problems for Helmholtz equations with Cauchy boundary conditions \cite{colton, isakov, hrycak}. In particular \cite{isakov, hrycak} show that for problems such as ours, the numerical stability of the inverse problem increases with the wavenumber $k$. 

When hitting a resonance, the field inside \eqref{eqn:potentials} the raceway varies significantly with small changes of the boundary data. This occurs for the forward problem and interferes in using the prior method. To determine this precisely, we can turn to the modal system \eqref{eqn:m_for} or \eqref{eqn:m_inv} and check if the matrix $\mathbf{M}_n$ is well conditioned, with one example shown in \Cref{fig:condition-heatmaps}.   

Due to the diffraction limit \cite{maznev2017upholding}, for any fixed frequency there is a limited amount of information, or resolution that can be extracted. 
The maximum spatial resolution that can be recovered from the boundary is given by the largest mode number $n$ used in the expansions \eqref{eqn:potentials}. When fixing $\omega$, the problem becomes more ill-posed as $n$ grows larger.


The condition number of $\mathbf M_n$ depends on the material parameters and geometry. In \Cref{sec:app2} we deduce an approximation to determine when the system is well posed, but this does not capture all the details shown in \Cref{fig:condition-heatmaps}. 
For instance, for the raceway in \Cref{tab:params} we can see from \Cref{fig:condition-heatmaps} that both $\mathbf{M}_n^\for$ and $\mathbf{M}_n^\inv$ can only be well conditioned if, approximately:
\begin{align} \label{eqn:M for well conditioned}
& |k_p| r_1 > |n|,
\end{align}
although there are many frequencies and modes $n$ that are ill posed for $\mathbf{M}_n^\for$ inside this region. Specifically, the condition number cond$\, \mathbf M_n^\for$ shows dark lines where the condition number is high. These indicate that the system is close to resonance, as small values of the boundary data lead to large values of the field. These dark lines depend on both $r_2$ and $r_1$, however we note that: when the ratio $r_2 / r_1$ gets closer to 1 the lines move to higher frequencies but get thicker, and when $r_2 / r_1$ gets larger, more and more lines move from high frequencies to lower frequencies, but get thinner and thinner. 

When the modal system is well conditioned, then the mode number $n$ can be measured when using $2n$ sensors, as illustrated in the example in \Cref{sec:example transient}. This is a rather high demand on sensors to reach a reasonable resolution. The number of sensors needed for roller bearings, developed in \Cref{sec:roller bearings}, is very different as we discuss below.

\begin{table}[!h]
\centering
\begin{tabular}{c|cc }
Parameter   & value & description \\
\hline
$r_1$  &    $1.0$\,m        &    \text{inner radius}
\\
$r_2$  &    $1.1$\,m        &    \text{outer radius}
\\
$c_p$      &    $5000$\,m/s         &    \text{pressure speed}  \\
 $c_s$  &  $3500$\,m/s  &     \text{shear speed}        \\
$\rho$  &    $7000$\,kg/m${}^3$        &  \text{mass density}   
\end{tabular}
\caption{the parameter values that approximate a steel raceway. These parameters are used for most numerical examples. \label{tab:params}}
\end{table}

\subsection{Roller bearings at constant speed}
\label{sec:measuring rollers}

Here our focus is to learn the Fourier coefficients of the loading profile $c_n$, shown in \eqref{eqn:loading profile fourier series}. In this section we explain an important lesson from the mathematics: the higher the rotation speed of the rollers $\Omega$, the better conditioned the modal system becomes, and the more coefficients $c_n$ it is possible to measure. Specifically, the lowest order coefficients $c_0$, $c_{-1}$, $c_{1}$, $c_{-2}$, $\ldots$, become inaccessibly if the speed $\Omega$ is too slow. 

Let us consider an example of a steel raceway with the properties shown in \Cref{tab:params}. To use the prior method, we need to invert the matrices $\mathbf M_n^\for$ as shown in \eqref{eqn:block prior boundary result}. This inversion is only stable when \eqref{eqn:M for well conditioned} holds. If we substitute the angular frequency $\omega_m = m Z \Omega$ from \eqref{eqn:contact points fourier} into \eqref{eqn:M for well conditioned} we obtain
\[
Z |m| C > |n|, \quad \text{where} \;\; C = \Omega r_1 / \alpha,
\]
note that the $\omega_m$ are the only frequencies available for constant rotation speed. Now the goal is to obtain the coefficients $c_{n-mZ}$ from \eqref{eqn:boundary to loading modes quasi static}. For clarity we define $\ell = n -m Z$ and substitute $n = \ell + m Z$ in the above, and manipulating, to reach the restriction:
\begin{equation} \label{ineq:loading}
    - Z (|m| C + m) < \ell < Z (|m| C - m).
\end{equation}

We can measure different frequencies $\omega_m$, which in turn implies we can choose different values for $m$. For each value $m$, the restriction \eqref{ineq:loading} determines which values for $\ell$ are possible to measure. Despite this liberty, if $C$ is small, then  \eqref{ineq:loading} will still significantly restrict all  possible values for $\ell$. For an example, assume that $C < 1$. For the raceway defined in \Cref{tab:params} we have that $C < 1$ when $\Omega < 10$ rev/min.

Let us consider the cases:
\begin{align}
    \hspace{2.5cm} & - Z (C + 1) < \ell < -Z (1 - C),  &\text{for}\;\; m = 1. \hspace{2.5cm}
    \\
    \hspace{2.5cm} & Z (1 - C) < \ell < Z (1 + C), &\text{for}\;\; m = -1, \hspace{2.5cm}
\end{align}
If $C = 0.5$ then the first and second inequality above would read $-1.5 Z < \ell < -0.5 Z$ and $0.5 Z < \ell < 1.5 Z$ respectively, which together imply that $|\ell| > 0.5 Z$. Larger values for $|m|$ would lead to restrictions where $|\ell|$ has to be larger. The number of rollers $Z$ can be anything larger than 10, so that $|\ell| > 0.5 Z$ would become $|\ell| > 5$. In other words, the loading coefficients $c_\ell$ for $\ell = -4,-3, \ldots ,4$ could not be reliably measured. 

The parameter $C$ can only increase, for one fixed raceway, when the speed of rotation $\Omega$ increases. With this increased speed, more modes of the loading become available to measure by measuring elastic waves. One way to interpret this is in terms of the static limit.  



\subsection{Static vs dynamic regimes}
\label{sec:static vs dynamic}

In the previous section we learned that if the rollers spin too slowly then the lowest order modes of the loading $|\ell|$ can be not be robustly measured. This is because as $\Omega$ slows down, we approach the static limit. A simple way to check if we are approaching the static limit is to compare the elastic wave speed with the speed of rotation of the rollers. That is, the ratio:
\begin{equation}
    \frac{\Omega r_1}{c_p} \;\; = \;\; \text{the roller to wave speed ratio}.
\end{equation}
If the above is very small, then the rollers are almost standing still relative to the wave speed, and therefore the solution could be calculated by using static stress balance, which is known to be ill-posed \cite{MARTIN1995825,MARIN2001783,marin-2002}, and the equations for the potentials \eqref{eqn:potentials} tend to Laplace equations which are also ill-posed\cite{isakov,kaipio2006statistical}. As discussed in the previous section, for the steel bearing in \Cref{tab:params} having $c = 1$ implies that $\Omega \approx 1.05\,$rad/s, which substituted into the ratio above leads to $2 \times 10^{-4}$. 
In conclusion, to predict the complete load due to the rollers (rotating at a constant speed) becomes well-posed if the rotation speed $\Omega$ is large enough. In practice, there are several ways around this limitation, as we explain next.

\noindent \textbf{Localised defects and forces.} Some important goals do not require a complete measurement of the loading through the rollers. An example of this is to detect a localised defect on, or near, the boundary in contact with the rollers. In this case, the Fourier coefficients $c_\ell$, of the loading profile,  for larger $\ell$ will be significant. These can be measured as shown in \Cref{sec:example localised defect for rollers}. Here we explain why this is possible in terms of algebra.

Suppose we are using a low frequencies $\omega_m$ which, due to the diffraction limit, implies we can only measure small values of $n$ of the boundary conditions $f_n$. Turning to \eqref{eqn:boundary to loading modes quasi static}, and setting as an example $n=0$, we could measure the coefficients $c_{-m Z}$ of the loading profile, where $m Z$ are high modes (as $Z > 10$)  which are related to locating defects, as shown in  \Cref{sec:example localised defect for rollers}.

\section{Examples} \label{sec:examples}
A large number of scenarios are rigorously tested in 
the folder \texttt{test} of the package \href{https://github.com/JuliaWaveScattering/ElasticWaves.jl/tree/v0.1.0}{\texttt{ElasticWaves.jl}} \cite{gower2024elasticwaves} where \href{https://github.com/JuliaWaveScattering/MultipleScattering.jl}{\texttt{MultipleScattering.jl}} \cite{gower2020multiplescattering} was also used. Below we show a few examples of methods developed in this paper to both validate and illustrate our method. 

In \Cref{sec:example transient} we show the modes for both the forward and inverse problem and explain where they are ill-posed. \Cref{sec:ex-localised force} shows an example of generating, and predicting, a localised for on the inner raceway. There we learn that many sensors are needed to accurately predict a localised mode if we make no assumptions about the forces. If we assume there are rollers travelling at a constant speed, then as shown in \Cref{sec:example-loading-profile} we can greatly reduce the number of sensors. Finally, \Cref{sec:example localised defect for rollers} shows an example where vibrations are due to rollers hitting a localised defect, and what can be recovered using the inverse system..

\subsection{The forward and inverse modal systems} \label{sec:example transient}

In the first sections of the paper we introduced the forward and inverse modal systems which are shown in \eqref{eqn:m_for} and \eqref{eqn:m_inv}. \Cref{fig:condition-heatmaps} shows where these systems likely lead to stable solutions. Here we provide examples that the inverse problem truly recovers the boundary conditions of the forward problem. We start with a sweep over all modes and frequencies.

\noindent \textbf{Boundary conditions.} For every mode $n$ and frequency $\omega$, let use choose the boundary conditions:
\[
\vec f_n = [1, 1, 0, 0]^{\mathrm{T}},
\]
for the forward problem to immitate some forcing on the inner face of the raceway. We then add a uniform random 2\% error and solve $\mathbf M_n^\for \vec a_n = \vec f_n$ for $\vec a_n$. Then, to setup the inverse problem, we substitute $\vec a_n$ in \eqref{eqn:potentials} and from these calculate the boundary data $\vec u_n$, the traction and displacement on the outer boundary. We then add 2\% error to $\vec u_n$ and solve $\mathbf M_n^\inv \vec a_n = \vec u_n$ for $\vec a_n$, and finally use this $\vec a_n$ to predict $\vec \tau_1$, the traction for $r  = r_1$. 

\begin{figure}
    \centering
    \includegraphics[width = 0.5\linewidth]{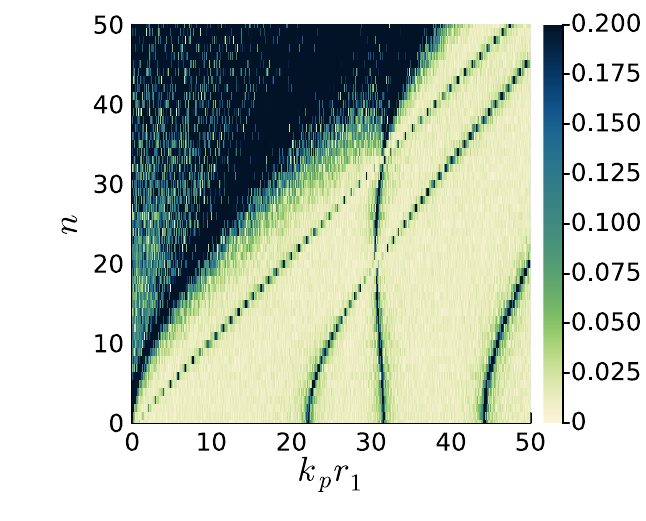} \hspace{-0.5cm}
    \includegraphics[width = 0.5\linewidth]{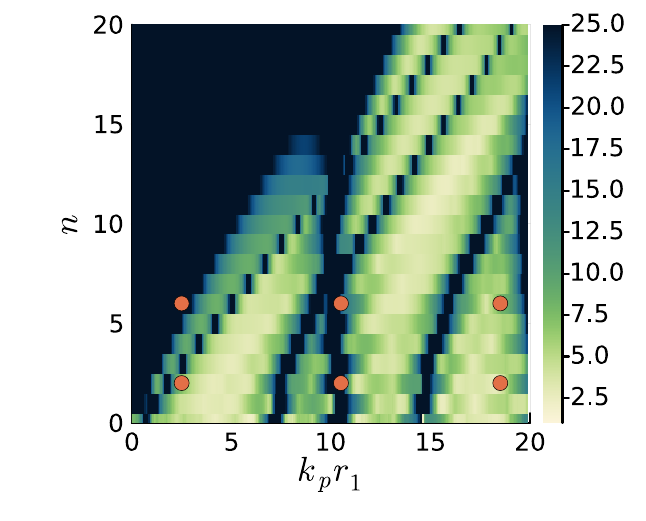} 
    \vspace{-0.52cm}
    
    \caption{On the left is the error in the traction $\vec \tau_1$, shown by \eqref{eqn:traction-error}, with $\vec \tau_1$ predicted by the inverse problem after adding 2\% error to all boundary data. \Cref{tab:params} shows the parameters used. The right shows the condition number of $\mathbf M_n^\for$ but for a thicker raceway: $r_1 = 1.0$ and $r_2 = 1.3$, which is easier to visualise the modes, where the modes are shown in \Cref{fig:modes} for the parameters of the orange spots. }
    \label{fig:condition-error}
\end{figure}

\noindent \textbf{The heatmap of error.} The error given by 
\begin{equation}\label{eqn:traction-error}
 \text{error} = |\vec \tau_1 - [1,1]^{\mathrm{T}}| / \sqrt{2}, 
\end{equation}
and is shown as a heatmap over all modes and frequencies in \Cref{fig:condition-error} (the image on the left). We can see that most of the heatmap has an error of around $2\%$, meaning that most modes and frequencies lead to a well posed problem.

\begin{figure}
    \centering
    \includegraphics[width=0.88\linewidth]{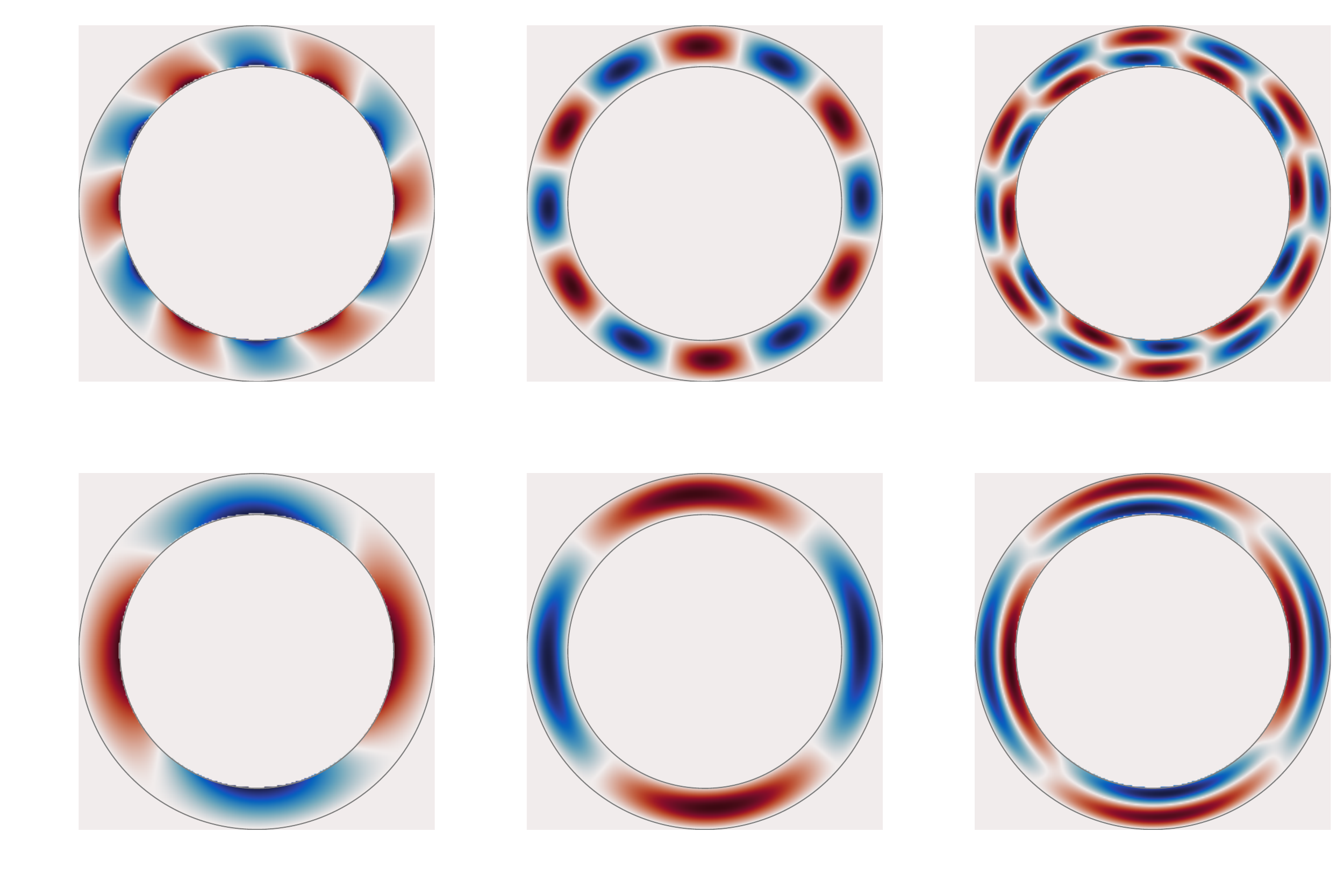}
    \vspace{-0.6cm}
    
    \caption{Above shows the real part of the pressure field from solving the forward system \eqref{eqn:m_for} for the modes $n = 2$ or $6$, and $k_p r_1 = 2.5, 10.5,$ or $18.5$. Red (blue) is positive (negative) pressure, while white is no pressure. The colour scaling is different for each mode, but the outer boundary is always traction-free so it is white. The outer radius $r_2 = 1.3$ which is easier to visualise then $r_1 = 1.1$. The chosen modes are shown as an orange scatter on \Cref{fig:condition-error}. Middle plots show resonance so large field inside for small boundary data.}
    \label{fig:modes}
\end{figure}

\noindent \textbf{Visualising the modes.} As explained in \Cref{sec:what is measureable}, the main causes that increase the error are 1) the diffraction limit and 2) resonant modes.  To help visualise, we plot some modes in \Cref{fig:modes}, with the mode number $n$ and wavenumber $k_p r_1$ of these modes shown by the orange spots in the heatmap on the right of \Cref{fig:condition-error}. We have chosen to use a thicker raceway with $r_1 = 1.0$ and $r_2 = 1.3$, but with the same material properties in \Cref{tab:params}, to better visualise the modes. The condition numbers of the forward problem for the thicker raceway is similar to the thinner raceway and is shown on the right of \Cref{fig:condition-error}.

\noindent \textbf{The diffraction limit modes.} The top and bottom left modes in Figure 1 are close to the diffraction limit, meaning the potentials approximately obey Laplace's equation. Notably, solutions to Laplace's equation are also solutions to a diffusion equation, where any source smoothly dissipates as it moves away from its origin. Solutions to the Laplacian are known to be ill posed and are essentially the same as the diffraction limit \cite{isakov,kaipio2006statistical}.

\noindent \textbf{Near resonant modes.} The two images in the middle column of \Cref{fig:modes} illustrate near resonant modes. That is the pressure on both boundaries is near zero, while the pressure away from the boundaries grows. This is why small errors in the boundaries lead to large errors in the fields for these modes. 

\noindent \textbf{Well-posed modes.} The top and bottom right modes shown in \Cref{fig:modes} are well posed, as both phase information is still present, and the pressure is not small on all boundaries.

\subsection{A localised force on the boundary}
\label{sec:ex-localised force}
In the previous section we saw that the inverse problem in general works, when outside of resonance of the low frequency limit. Here we show an example where although the inverse problem is well posed, you would need many Fourier modes to converge, and therefore many sensors. The material parameters and dimensions of the bearing used in these simulations are laid out in \Cref{tab:params}. This section acts as motivation for using the prior method developed in \Cref{sec:prior method}.

\noindent \textbf{Boundary conditions and method.} Consider a sharp Gaussian force applied to the inner boundary given by 
\begin{equation}\label{eqn:trans-time}
    f(\theta) = \frac{1}{\sigma \sqrt{2\pi}} \e^{-\frac{(\theta-\pi)^2}{2\sigma^2}},
\end{equation}
where $\sigma = 0.1$. 
For the forward model we use the boundary conditions
\begin{equation*}
    \vec \tau^1 = f(\theta)\hat{\vec r} \quad \text{and} \quad \vec \tau^2 = \vec 0,
\end{equation*}
which for one fixed frequency $\omega$ leads to the modal system 
\begin{equation}\label{eqn:modes-gauss}
    \mathbf{M}^{\for}_n \boldsymbol a_n = 
        [f_n, 0, 0, 0]^T,
\end{equation}
where $f_n$ is the $n$th coefficient of the Fourier series expansion of the forcing \eqref{eqn:trans-time}. A large number of coefficients $f_n$ are needed to accurately represent $f(\theta)$, which is why this example will need many sensors to obtain a good resolution. Solving \Cref{eqn:modes-gauss} for each mode $n$ gives a solution to the forward problem.

\begin{figure}[h!]
    \centering
    \includegraphics[width=0.45\linewidth]{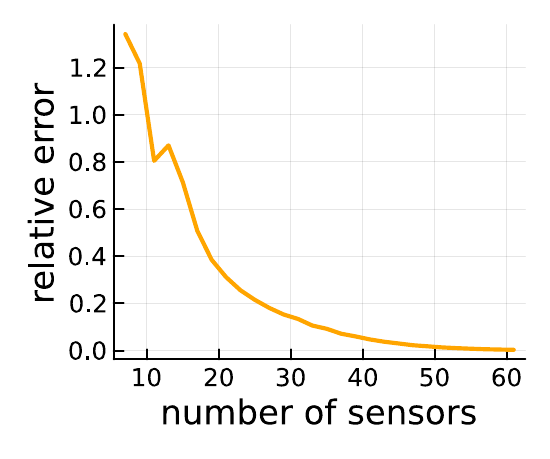}
    \includegraphics[width=0.51\linewidth]{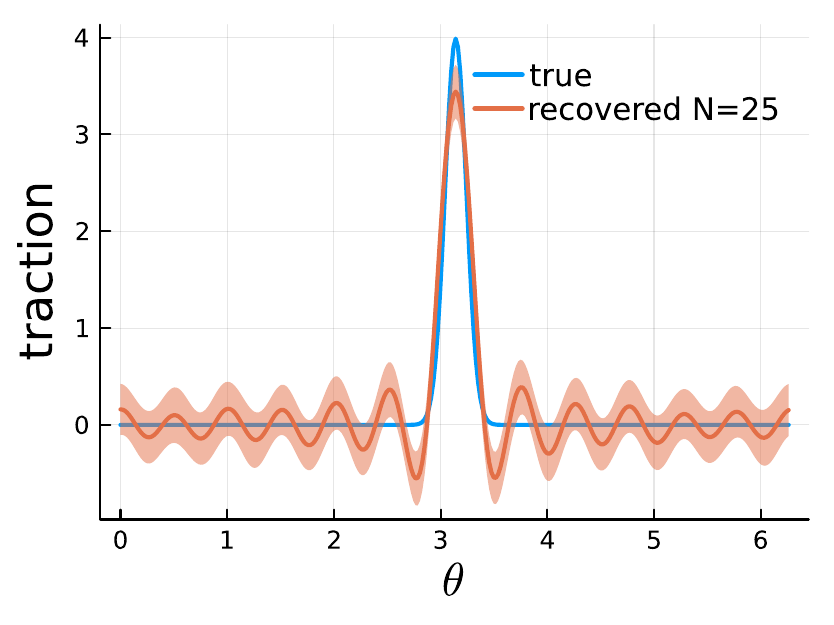}
    \caption{The left shows the convergence of solving the inverse problem, with 1\% error added to its boundary conditions, to recover the force shown on the right. As the number of sensors increases so does the number of Fourier modes $n$ that can be recovered. The right shows the forcing $f(\theta)$ given by \eqref{eqn:trans-time} compared with the mean predicted force from the inverse problem for $k_p r_1 = 65$, using 25 sensors, and with a $10$ standard deviation ribbon to show the uncertainty.}
    \label{fig:trans-inv}
\end{figure}

Just as before, to setup the inverse problem, we use the forward problem to predict the displacement $\vec u_n$ on the outer boundary $r = r_2$, and then add 1\% error to $\vec u_n$ and then solve the inverse modal problem \eqref{eqn:m_inv} for $\vec a_n$. With $\vec a_n$ we then predict the traction $\vec \tau_1$ and compare it with the true traction. The results for the frequency $k_p r_1 = 65$ are shown in \Cref{fig:trans-inv} where the ribbon is $10$ the standard deviation from solving this problem many times each with a different error added. 

\noindent \textbf{Results.} From the thickness of the ribbon in the plot on the right of \Cref{fig:trans-inv} we can see that the problem is well posed for $k_p r_1 = 65$. This relatively high frequency avoids most resonances, and allows us to recover very high Fourier modes by avoiding the diffraction limit. However, the number of sensors needed to reach a relative error less than 20\% is around 25, as can be seen from the graph on the left of \Cref{fig:trans-inv}.  


Clearly a Fourier series representation $f(\theta) = \sum_n f_n \ee^{\ii n \theta}$ is not the best choice for a localised force. However, by making assumptions about what led to the force, or more generally the boundary conditions, we can use many other representations. For example, we can assume the traction is due to contact with rollers bearings rotating at a constant speed, as we do in the next section.


\subsection{Recovering the loading profile for rollers} \label{sec:example-loading-profile}

In this section we show an example of predicting the loading on the bearings by using the methods developed in \Cref{sec:static-loading-profile}. That is, we assume that the rollers are rotating at a constant speed.

\noindent \textbf{Stribeck boundary condition.} To create a realistic boundary conditions for the inverse problem we use the Stribeck equation for the loading profile of roller bearings  \cite{harris2001rolling}, it is given by
\begin{equation} \label{eqn:stribeck}
    L(\theta)= L_0 \left( 1- \frac{1}{2\epsilon}(1- \cos{\theta}) \right)^{10/9},
\end{equation}
where $\epsilon$ is called the load distribution factor. This parameter determines the loading zone, that is, the region where the load is being applied. For a radial loading, it is related to the loading region through:
\begin{equation}
    \epsilon=\frac{1}{2} \left( 1 - \cos (\varphi/2)\right)
\end{equation}
where $\varphi$ is the angular extent of the loading zone.

For the numerical simulations, we used $L_0=1$ and $\epsilon=0.5$, which implies a loading zone of angular length of $\varphi= \pi$. \Cref{fig:true and prior stress} shows this Stribeck loading profile. The other parameters used for this section are shown in \Cref{tab:loading-profile}, where we use a thicker raceway with thickness 1m to make the plots below easier to see.

\begin{table}[!h]
\centering
\begin{tabular}{c|cc }
Parameter   & value & description \\
\hline
$r_1$  &    $2.5$\,m        &    \text{inner radius}
\\
$r_2$  &    $3.5$\,m        &    \text{outer radius}
\\
$c_p$      &    $5000$\,m/s         &    \text{pressure speed}  \\
 $c_s$  &  $3500$\,m/s  &     \text{shear speed}        \\
$\rho$  &    $7800$\,kg/m${}^3$        &  \text{mass density}   
\\
$\Omega$  &    $2000$\,rpm        &   \text{rotation speed}
\end{tabular}
\caption{Parameter values used for numerical simulations in \Cref{sec:example-loading-profile}.}
\label{tab:loading-profile}
\end{table}

\noindent \textbf{Data from the forward problem.} Like the previous sections, we create the boundary data for the inverse problem, represented by $\vec y^\inv$ in  \eqref{eqn:block modal inverse}, by solving the forward problem. In this section, for the forward problem we used the loading \eqref{eqn:stribeck}, from which the $c_n$ coefficients in \eqref{eqn:loading profile fourier series} can be calculated, which then lead to the Fourier modes $f_n$ shown in \eqref{eqn:roller loading to coefficients} which we use for the boundary conditions of the forward problem. Again we assume the outer boundary $r = r_2$ is traction free with $\vec \tau_1 = \vec 0$. 

After solving the forward problem, we can then calculate $\vec y^\inv$ from \eqref{eqn:bounday block discrete} for a chosen number of sensors, where $\vec a$ is given by solving the forward problem, and $\mathbf E$ is composed of the modal matrices for the inverse problem. 

\begin{figure}
    \centering
    \includegraphics[width=0.38\linewidth]{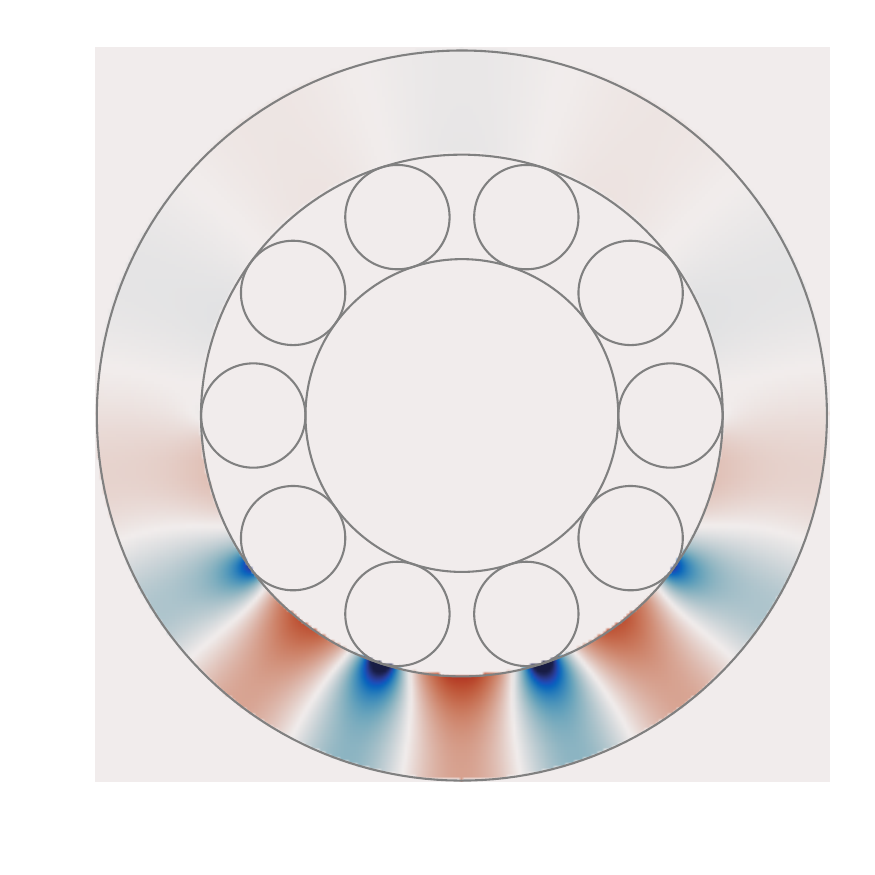}
    \includegraphics[width=0.38\linewidth]{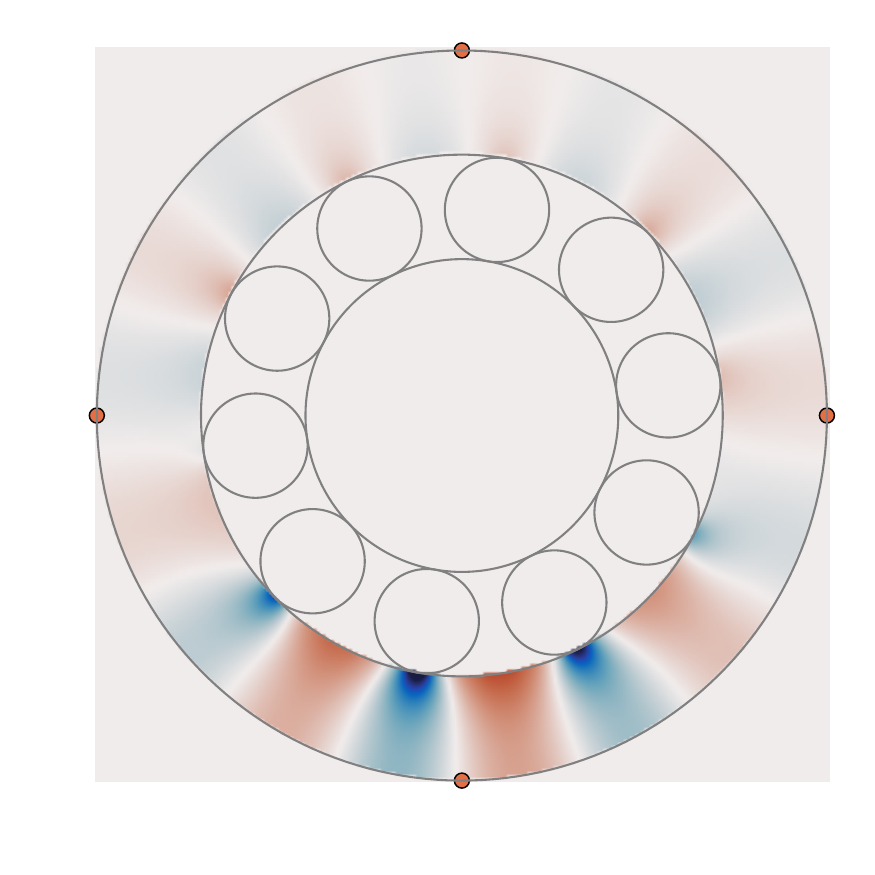}
    \caption{True radial displacement on the left for one snapshot in time when using the Stribeck equation \eqref{eqn:stribeck} and the properties in \Cref{tab:loading-profile}. The snapshot in time is a result of taking a Fourier transform over all frequencies. The right shows the predicted radial displacement when using only 4 sensors shown as orange spots. The sensors measure displacement, and the outer boundary is stress free. We can see that despite being stress free on the boundary $r = r_2$, the displacement is not zero there. The recovery is not perfect as 4\% noise is added, and limit sensors imply limited Fourier modes are recovered.}
    \label{fig:true and prior displace}
\end{figure}

\begin{figure}[h!]
    \centering
    \includegraphics[width=0.65\linewidth]{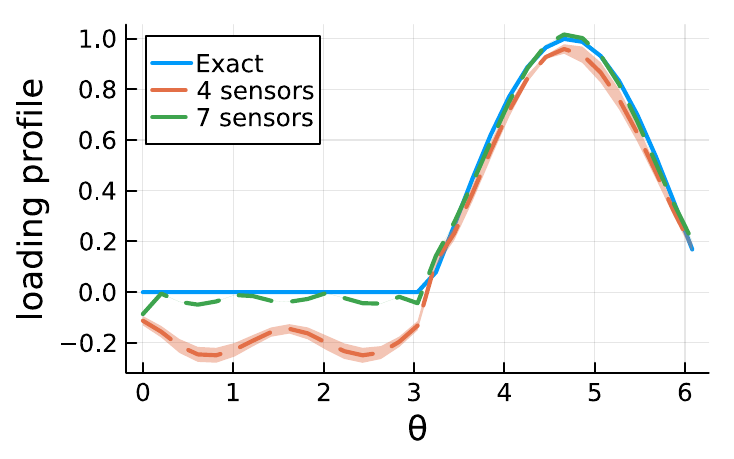}
    \caption{The blue curve shows the Stribeck equation for the loading of rollers given in \eqref{eqn:stribeck}, the orange is prior method with 4 sensors and modes -2:2 and 4\% added error to boundary data, other than 20\% error caused from lacking Fourier mode. Also added 4\% white noise error on top. Green is also modes -2:2 and 4\% added error, but with 7 sensors able to resolve and ignore higher Fourier modes. }
    \label{fig:stribeck and predicted}
\end{figure}


\noindent \textbf{Results.} To solve the inverse problem using the prior method we can follow the steps shown at the end of \Cref{sec:static-loading-profile}. The result is that at least 4 sensors are needed to recover the loading profile, when adding 4\% noise, as shown by the  \Cref{fig:true and prior stress} in the introduction, which shows the predicted pressure distribution, and \Cref{fig:true and prior displace} which shows the predicted displacement. \Cref{fig:stribeck and predicted} shows a more quantitative view with just the predicted loading profile $L(\theta)$ when using 4 and 7 sensors, shown against the exact loading profile used. When using 4 or 7 sensor we are only trying to recover the Fourier coefficients $c_{-2}, c_{-1}, c_{0}, c_{1}$, and $c_{2}$. When using 4 sensors there is a reasonably large error because the higher Fourier coefficients, which are ignored, make a substantial contribution to the Stribeck equation shown in \Cref{fig:stribeck and predicted}. With only 4 sensors, the ignored higher Fourier coefficients are treated liked an add error (20\%), which explains the error in recovering the coefficients $c_{-2}, c_{-1}, c_{0}, c_{1}$, and $c_{2}$. When using 7 sensors, we are able to differentiate the modes associate with $c_{-3},  c_3$ from the lower modes.

\subsection{Localised defect in a roller bearing}
\label{sec:example localised defect for rollers}
\begin{figure}[h]
    \centering
 %
\begin{tikzpicture}[rotate=0,scale=0.74]
    \coordinate (O) at (0,0);
    \draw[fill=gray!30] (O) circle (2.8);
    \draw[fill=gray!80] (O) circle (2);
    \draw[fill=gray!30] (O) circle (1.2);
    \draw[fill=gray!40] (O) circle (0.6);
    \draw[fill=gray!30] (1.6,0) circle (0.4);
    \draw[fill=gray!30] (-1.6,0) circle (0.4);
    \draw[fill=gray!30] (0,-1.6) circle (0.4);
    \draw[fill=gray!30] (0,1.6) circle (0.4);
    \draw[fill=gray!30] (1.13137,1.13137) circle (0.4);
    \draw[fill=gray!30] (-1.13137,-1.13137) circle (0.4);
    \draw[fill=gray!30] (1.13137,-1.13137) circle (0.4);
    \draw[fill=gray!30] (-1.13137,1.13137) circle (0.4);
    \draw[fill=gray!80] (2.8,-0.25) rectangle (3,0.25);
    \draw[fill=gray!80, rotate=45] (2.8,-0.25) rectangle (3,0.25);
    \draw[fill=gray!80, rotate = -45] (2.8,-0.25) rectangle (3,0.25);
    \draw[fill=gray!80, rotate=90] (2.8,-0.25) rectangle (3,0.25);
    \draw[fill=gray!80, rotate = -90] (2.8,-0.25) rectangle (3,0.25);
    \draw[fill=gray!80, rotate=225] (2.8,-0.25) rectangle (3,0.25);
    \draw[fill=gray!80, rotate = -225] (2.8,-0.25) rectangle (3,0.25);
    \draw[fill=gray!80, rotate=180] (2.8,-0.25) rectangle (3,0.25);
 \filldraw[red!40!gray!30,shift={(2,0)}, rotate=-70] (0,0) - - (0.75,0) arc (0:140:0.75) - - cycle;
  \fill[red] (2,0) circle (\p);
  \foreach \i in {1,...,3}{
    \draw[red!80,ultra thick, shift={(2,0)}] (70:\r*\i) arc(70:-70:\r*\i);
  }
  \node[inner sep=0pt] at (-8.5,0)
    {\includegraphics[width=0.48\linewidth]{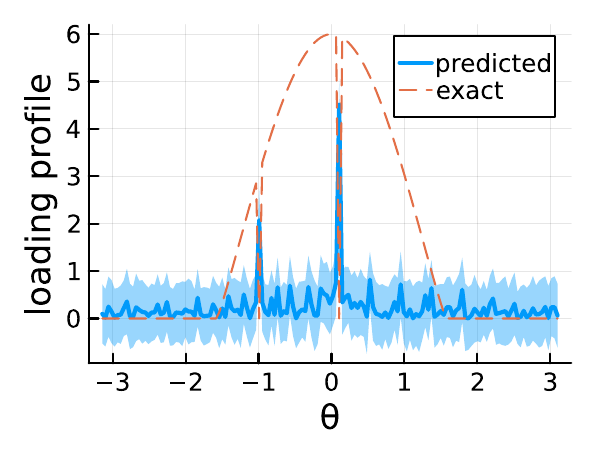}};
\end{tikzpicture}
\vspace{-0.35cm}

\caption{
The right is an illustration of elastic waves being emitted when a roller hits a defect on the raceway. 
The left shows the loading profile (orange and dashed) with two sharp drops in pressure due to the presence of two defects on the inner boundary. The blue curves show the absolute value of the predicted loading profile when measuring the Fourier coefficients of the boundary data $\vec u_n$ for $n = -6,-5,\ldots, 6$ where 2\% error was added. The rollers are rotating at a rate of $\Omega = 120$ rpm and the properties used for the raceway are shown in \Cref{tab:params}. 
}
\label{fig:bearing_measure}
\end{figure}

As our final example, we consider a a localised defect on, or near, the boundary in contact with the rollers. A schematic is shown on the right of \Cref{fig:bearing_measure}.

\noindent \textbf{Defects and slow rotation.} In the previous example we showed how a smooth loading profile can be predicted with only a few sensors by using the prior method, together with a Fourier series expansion of the loading profile, as shown in \Cref{sec:static-loading-profile}. To predict the loading profile, the rollers need to rotate fast enough, as discussed in \Cref{sec:static vs dynamic}. If the rollers are rotating slower, then we can only predict the higher Fourier coefficients $c_\ell$ for larger $\ell$, which are associated to localised defects, as we illustrate in this example.

\noindent \textbf{Use the inverse modal system.} Consider the inside raceway with two localised defects which we assume leads to a loading profile shown by the orange dashed curve on the left of \Cref{fig:bearing_measure}. 
This time, rewriting the loading profile in terms of a Fourier series, as done in \Cref{sec:static-loading-profile}, does not help because the Fourier series will converge very slowly. So instead, just to illustrate, we just directly solve the system \eqref{eqn:bounday block discrete} using the inverse modal system \eqref{eqn:m_inv} and boundary data $\vec y^\inv$ solely from the outer boundary. 



We return to using the parameters in \Cref{tab:params}, as these more closely match real applications, but use now a slower rotation speed of the rollers $\Omega = 120\,$rpm. As discussed in \Cref{sec:measuring rollers}, for slow rotation speeds there are restrictions on which of the Fourier coefficients $c_\ell$, of the loading profile, can be measured. 

\noindent \textbf{What can be measured.} Each frequency $\omega_m$ gives access to a range of values for $\ell$. For this example we use the frequencies $m = 1,2,\ldots 5$ and for each solve \eqref{eqn:bounday block discrete} for $\vec a$, and then predict the inner traction $\vec \tau_1$. From $\vec \tau_1$, and depending on the choice of $m$, we then estimate some of the coefficients $c_\ell$ of the loading profile by using \eqref{eqn:roller loading to coefficients}. By combining all the predicted coefficients $c_\ell$ from all five frequencies $\omega_m$ we then predict the absolute value loading profile shown by the blue curve in \Cref{fig:bearing_measure}.  We have shown the absolute value for visual clarity, as we can more clearly see that the blue and orange spikes match up perfectly.

Clearly, \Cref{fig:bearing_measure} shows that we are able to both locate the defects, and determine their magnitude, at least in terms of the pressure difference, by directly solving the inverse problem.

\noindent \textbf{Envelope analysis.} There is a method commonly used to detect localised defects called \emph{envelope analysis} \cite{randall2001relationship}.   Here we only make a brief comment on how this method is connected to the work in this paper. 

For the example in this section, we are not able to recover the first modes $c_0$, $c_{-1}$, or $c_{1}$ of the loading profile. However, these modes are not small, and therefore do make a contribution to our boundary data. They in fact act like an error term. For example, for the frequency $\omega_m$, the coefficient $c_0$ contributes to $f_{mZ}(\omega_m)$ of the boundary data as shown in \eqref{eqn:roller loading to coefficients}. If we have less than $m \times Z$ sensors, and attempt to calculate a Fourier series of the loading data $y(\theta,\omega_m)$, then the mode of $f_{mZ}(\omega_m)$ will be mixed in with the other modes $f_{n}(\omega_m)$, for $|n| < m \times Z$, and lead to errors for these modes. This error could be avoided if the function $y(\theta,\omega_m)$ was first smoothed in $\theta$ before calculating the Fourier modes, as the smoothening would remove the higher modes such as $f_{mZ}(\omega_m)$. We believe this can be linked with the smoothening in time used in Envelope Analysis, though this deserves a more lengthy analysis elsewhere.

\section{Conclusions}

In this paper we have shown how to model elastic waves confined within a hollow thick walled cylinder with symmetry along the axis. As discussed in \Cref{sec:maths}, the dynamics of these waves captures the dominant vibrations within a raceway. By deriving simple systems for the modes, we provided tools to better understand and quickly solve for these elastic waves. A detailed outline of the paper's content is given at the end of \Cref{sec:intro}.

\noindent \textbf{Results.} The main results are how to: 1) model waves, 2) use prior assumptions about the boundary conditions, and 3) determine what it is, and is not, possible to predict the traction or displacement within the raceway. Notably, in \Cref{sec:what is measureable}, we show that solving for the elastic waves becomes ill-posed when hitting a resonant frequency or near the diffraction limit. These results hold for any transfer path of the signal. For roller bearings we demonstrate that if the rotation speed is slow, then it is only possible to predict localised contact forces. Extended, or smooth, contact forces lead to ill-posed problems for elastic waves.

\noindent \textbf{Modelling - raceways.} Our models lay the foundation for many future avenues. For instance, instead of considering waves which are just confined in the raceway, as shown by \Cref{fig:modes}, the boundary conditions can be adjusted to let waves leak out towards the rollers or the oil. Additionally, the raceway's bolted supports can be incorporated in the boundary conditions by assuming that waves dissipate through these bolts and are not reflected back into the raceway. 

\noindent \textbf{Modelling - bearings.} It is also possible to extend the models to consider bearings that do not have axial symmetry, such as ball bearings or steeply inclined tapered roller bearings. Further extensions could account for the slip and slide of roller bearings \cite{randall2011rolling}  which lead to transient waves with a high-frequency content. Incorporating these phenomena into the models as more elaborate priors would lead to more accurate predictions.

\noindent \textbf{Inverse problem - uncertainty.} To develop robust predictions, and methods to determine defects, uncertainty needs to be accounted for \cite{jones2021bayesian}. A first step in this direction is to consider the boundary data to be samples of a distribution, and also to consider that priors, as discussed in \Cref{sec:linear priors}, are also distributions. This would help properly account for roller bearings slipping, or fluid interaction in journal bearings.

\noindent \textbf{Detect localised defects.} In industrial applications, it is common to have very few sensors per bearing, which limits the diagnostic methods available. With only a few sensors, defect detection often relies on  monitoring the amplitude of specific frequencies, such as the ball pass frequencies, to identify defects. Can we rely on these methods? The models in this paper provide a clear path to address this question: imagine continually measuring the vibration of a bearing. When a change occurs, we can assume it is due to a localised defect with an unknown position and size. By adopting a Bayesian approach, we can then estimate the defect size by marginalising over its possible positions. This approach would clarify how robust current methods are and lead to more reliable, physics-based diagnostic methods to detect defects. 

\appendix

\section{Entries of $\mathbf{M}_n^{\text{for}}$ and $\mathbf{M}_n^{\text{inv}}$}\label{sec:app3}
Equation \eqref{eqn:m_for} may be written as
\begin{align*}
    \begin{pmatrix}
        P^{(n)}_1(r_1)&P^{(n)}_2(r_1)&P^{(n)}_3(r_1)&P^{(n)}_4(r_1)\\
        S^{(n)}_1(r_1)&S^{(n)}_2(r_1)&S^{(n)}_3(r_1)&S^{(n)}_4(r_1)\\
        P^{(n)}_1(r_2)&P^{(n)}_2(r_2)&P^{(n)}_3(r_2)&P^{(n)}_4(r_2)\\
        S^{(n)}_1(r_2)&S^{(n)}_2(r_2)&S^{(n)}_3(r_2)&S^{(n)}_4(r_2)
    \end{pmatrix}
    \begin{pmatrix}
        a_n\\
        b_n\\
        c_n\\
        d_n
    \end{pmatrix}
    =
    \begin{pmatrix}
        -p^1_n\\
        -s^1_n\\
        p^2_n\\
        s^2_n
    \end{pmatrix}
\end{align*}
where,
\begin{align*}
   P^{(n)}_1(r)&=-\frac{\rho}{r^2}\left(2 c_s^2 k_p r \besselj_{n-1}(k_pr) + \left(r^2 \omega^2 - 2c_s^2 n(n+1)\right)\besselj_{n}(k_pr)\right)\\
   P^{(n)}_2(r)&=-\frac{\rho}{r^2}\left(2 c_s^2 k_p r \hankelh^{(1)}_{n-1}(k_pr) + \left(r^2 \omega^2 - 2c_s^2 n(n+1)\right)\hankelh^{(1)}_{n}(k_pr)\right)\\
   P^{(n)}_3(r) &= \frac{\im\rho c_s^2 n}{r^2} \left(k_sr\besselj_{n-1}(k_sr)-2\besselj_{n}(k_sr)-k_sr \besselj_{n+1}(k_sr)\right)\\
   P^{(n)}_4(r)&= \frac{\im\rho c_s^2 n}{r^2} \left(k_sr\hankelh^{(1)}_{n-1}(k_sr)-2\hankelh^{(1)}_{n}(k_sr)-k_sr \hankelh^{(1)}_{n+1}(k_sr)\right)\\
   S^{(n)}_1(r)&=\frac{\im\rho c_s^2 n}{r^2} \left(k_pr\besselj_{n-1}(k_pr)-2\besselj_{n}(k_pr)-k_pr \besselj_{n+1}(k_pr)\right)\\
    S^{(n)}_2(r)&=\frac{\im\rho c_s^2 n}{r^2} \left(k_pr\hankelh^{(1)}_{n-1}(k_pr)-2\hankelh^{(1)}_{n}(k_pr)-k_pr \hankelh^{(1)}_{n+1}(k_pr)\right)\\
    S^{(n)}_3(r)&=\frac{\rho}{r^2}\left(2 c_s^2 k_s r \besselj_{n-1}(k_sr) + \left(r^2 \omega^2 - 2c_s^2 n(n+1)\right)\besselj_{n}(k_sr)\right)\\
    S^{(n)}_4(r)&=\frac{\rho}{r^2}\left(2 c_s^2 k_s r \hankelh^{(1)}_{n-1}(k_sr) + \left(r^2 \omega^2 - 2c_s^2 n(n+1)\right)\hankelh^{(1)}_{n}(k_sr)\right)
\end{align*}

Equation \eqref{eqn:m_inv} may be written as
\begin{align*}
    \begin{pmatrix}
        U^{(n)}_{r,1}(r_2)&U^{(n)}_{r,2}(r_2)&U^{(n)}_{r,3}(r_2)&U^{(n)}_{r,4}(r_2)\\
        U^{(n)}_{\theta,1}(r_2)&U^{(n)}_{\theta,2}(r_2)&U^{(n)}_{\theta,3}(r_2)&U^{(n)}_{\theta,4}(r_2)\\
        P^{(n)}_1(r_2)&P^{(n)}_2(r_2)&P^{(n)}_3(r_2)&P^{(n)}_4(r_2)\\
        S^{(n)}_1(r_2)&S^{(n)}_2(r_2)&S^{(n)}_3(r_2)&S^{(n)}_4(r_2)
    \end{pmatrix}
    \begin{pmatrix}
        a_n\\
        b_n\\
        c_n\\
        d_n
    \end{pmatrix}
    =
    \begin{pmatrix}
        u^{(r)}_n\\
        u^{(\theta)}_n\\
        p^2_n\\
        s^2_n
    \end{pmatrix}
\end{align*}
where,
\begin{align*}
    U^{(n)}_{r,1}(r) &= \frac{k_p}{2}\left(\besselj_{n-1}(k_pr) - \besselj_{n+1}(k_pr)\right), \quad 
    U^{(n)}_{r,2}(r) = \frac{k_p}{2}\left(\hankelh^{(1)}_{n-1}(k_pr) - \hankelh^{(1)}_{n+1}(k_pr)\right),
    \\
    U^{(n)}_{r,3}(r) &= \frac{\im n}{r} \besselj_n(k_sr), \;\; 
    U^{(n)}_{r,4}(r) = \frac{\im n}{r} \hankelh^{(1)}_n(k_sr), \;\;
    U^{(n)}_{\theta,1}(r) = \frac{\im n}{r} \besselj_n(k_pr), \;\;
    U^{(n)}_{\theta,2}(r) = \frac{\im n}{r} \hankelh^{(1)}_n(k_pr)
    \\
    U^{(n)}_{\theta,3}(r) &= \frac{-k_s}{2}\left(\besselj_{n-1}(k_sr) - \besselj_{n+1}(k_sr)\right), \quad 
    U^{(n)}_{\theta,4}(r) = \frac{-k_s}{2}\left(\hankelh^{(1)}_{n-1}(k_sr) - \hankelh^{(1)}_{n+1}(k_sr)\right).
\end{align*}

\section{Derivation of traction components}\label{sec:app1}
In this section we show how to derive equations \eqref{eqn:sig_rr} and \eqref{eqn:sig_rth}. We begin by specifying the form of the stress tensor. In our case, since the material we are considering is homogeneous and isotropic the desired form is given by equation \eqref{eqn:constitutive}.
Using \eqref{eqn:constitutive}, we obtain the following expressions for $\sigma_{rr}$ and $\sigma_{r\theta}$ it terms of the displacement $\vec u$
 \begin{equation}\label{eqn:sigrr and sigrth}
     \sigma_{rr} = \lambda \left(\frac{\partial u_r}{\partial r} + \frac{1}{r}\left(\frac{\partial u_{\theta}}{\partial \theta} + u_r \right) \right) + 2\mu \frac{\partial u_r}{\partial r}, \quad
     \sigma_{r\theta} = \frac{\mu}{r}\left(\frac{\partial u_{r}}{\partial \theta} + r \frac{\partial u_{\theta}}{\partial r} - u_{\theta} \right).
 \end{equation}
We will deduce both equations in turn, beginning with $\sigma_{rr}.$  Firstly, note that \eqref{eqn:sigrr and sigrth}${}_1$ may be rewritten as 
\begin{equation}
    \sigma_{rr} = (\lambda+2\mu) \nabla \cdot \boldsymbol u -  \frac{2\mu}{r}\left(\frac{\partial u_{\theta}}{\partial \theta} + u_r \right).
\end{equation}
From $\boldsymbol u=\nabla \phi + \nabla \times (\psi \hat{\vec z})$ we see that $\nabla \cdot \boldsymbol u = \lap \phi = -k_p^2 \phi$, hence
\begin{align*}
\sigma_{rr}  = -k_p^2(\lambda+2\mu)\phi-  \frac{2\mu}{r}\left(\frac{\partial u_{\theta}}{\partial \theta} + u_r \right).
\end{align*}
Substituting \eqref{eqn:helmdec} into \eqref{eqn:helm-phi} leads to
\begin{equation*}
    \frac{1}{r}\left(\frac{\partial u_{\theta}}{\partial \theta} + u_r\right) = -k_p^2 \phi - \frac{\partial^2 \phi}{\partial r^2} - \frac{\partial}{\partial r} \left(\frac{1}{r}\frac{\partial \psi}{\partial \theta} \right),
\end{equation*}
which itself substituting into $\sigma_{rr}$ above leads to
\begin{align}
\sigma_{rr} = -\lambda k_p^2 \phi + 2\mu \left(\frac{\partial^2 \phi}{\partial r^2} + \frac{\partial}{\partial r} \left(\frac{1}{r}\frac{\partial \psi}{\partial \theta} \right) \right).
\end{align}
Finally using \eqref{eqn:Lame parameters} we deduce  $\lambda = \rho(c_p^2 - 2c_s^2)$ which together with $\omega = c_p k_p$ substituted above leads to \eqref{eqn:sig_rr}.

To simplify $\sigma_{r\theta}$ in \eqref{eqn:sigrr and sigrth}${}_2$ first we rewrite it in the form
\begin{align*}
\sigma_{r\theta} = \mu \left(\nabla \times \boldsymbol u \right) \cdot \mathbf{\hat{z}} - \frac{2\mu}{r}\left(u_{\theta} - \frac{\partial u_r}{\partial \theta} \right).
\end{align*}
Now, from \eqref{eqn:helmdec} and \eqref{eqn:helm-phi}, we have $\left(\nabla \times \boldsymbol u \right) \cdot \mathbf{\hat{z}} = -\lap \psi = k_s^2 \psi$, which substituted above leads to
\begin{align*}
\sigma_{r\theta} = \mu k_s^2 \psi - \frac{2\mu}{r}\left(u_{\theta} - \frac{\partial u_r}{\partial \theta} \right).
\end{align*}
Using \eqref{eqn:helmdec} and \eqref{eqn:helm-phi} we find that 
\begin{equation*}
    \frac{1}{r}\left(u_{\theta} - \frac{\partial u_r}{\partial \theta} \right) =
    k_s^2\psi + \frac{\partial^2 \psi}{\partial r^2}-\frac{\partial}{\partial r} \left(\frac{1}{r}\frac{\partial \phi}{\partial \theta}\right),
\end{equation*}
which substituted into $\sigma_{r\theta}$ above leads to
\begin{align*}
\sigma_{r\theta} = -\mu k_s^2 \psi - 2\mu \left(\frac{\partial^2 \psi}{\partial r^2}-\frac{\partial}{\partial r} \left(\frac{1}{r}\frac{\partial \phi}{\partial \theta}\right) \right),
\end{align*}
which is the same as \eqref{eqn:sig_rth} after using $\mu = \rho c_s^2$ and $\omega = c_s k_s$.

\section{The Diffraction Limit}\label{sec:app2}

One cause of the modal system becoming ill-posed, as shown in \Cref{sec:what is measureable}, is due to the diffraction limit \cite{maznev2017upholding}. 
Below we provide an approximate formula based classical argument of the diffraction limit to easily determine when the system is ill-posed.

Instead of solving a boundary value problem, we consider a simpler case of determining the amplitude $A$ and $B$ of two point sources on the boundary $r = r_1$. These in sense imitate to forcing on the boundary. The field emitted by these sources is given by
\begin{equation}\label{eqn:src}
    \phi_{\text{src}} = \frac{\im}{4}\left(A\hankelh^{(1)}_0(k_p|\vec r - \vec r^{(1)}|)+B\hankelh^{(1)}_0(k_p|\vec r - \vec r^{(2)}|)\right),
\end{equation}
with $\vec r^{(1)}$ and $\vec r^{(2)}$ representing the position of the first and second point source respectively.

\begin{figure}[h]
    \centering
    \includegraphics[width=0.55\linewidth]{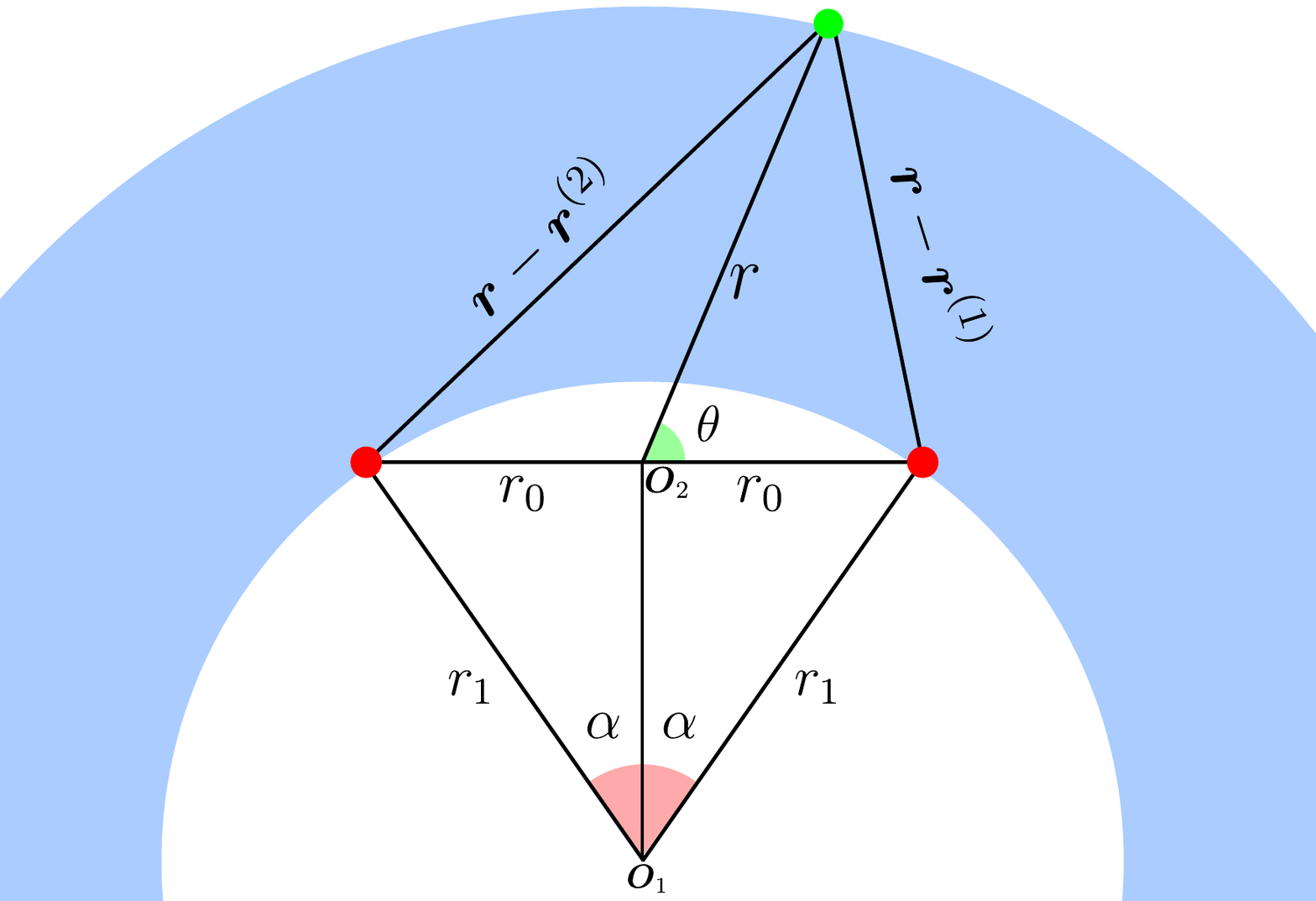}
    \caption{Illustration of Graff's addition theorem. $r_0$ is the distance from $\vec O_2$ to each source, $r$ is the distance from $\vec O_2$ to some observation point on the outer boundary, and $\theta$ is the angle of the observation point from the source at $\vec r^{(1)}$. The angles $\alpha$ are needed to relate these sources to Fourier modes later.}
    \label{fig:graff}
\end{figure}

Now we ask, can we tell the different between these two sources by measuring the field on boundary $r = r_2$? That is, can we distinguish between a source at $\vec r^{(1)}$ from $\vec r^{(2)}$? As we want to determine the maximum amount of information available we consider that we have access to the field everywhere on the boundary $r = r_2$. To achieve this it is convenient to use the origin $\vec O_2$, the midpoint of the chord connecting the two sources. Then 
we can use Graff's addition theorem \cite{abramowitz+stegun, napal2024effective} to rewrite \eqref{eqn:src} as
\begin{equation}\label{eqn:src_graff}
    \phi_{\text{src}} = \frac{\im}{4}\sum_n \left(A \besselj_{-n}(k_pr_0)\hankelh^{(1)}_n(k_pr)+B \besselj_{n}(k_pr_0)\hankelh^{(1)}_n(k_pr)\right)\e^{\im n\theta},
\end{equation}
where $r_0$ is the horizontal distance from $\vec O_2$ to each source, $r$ is the distance from $\vec O_2$ to some observation point on the outer boundary, and $\theta$ is the angle of the observation point from the source at $\vec r^{(1)}$, this is illustrated in \Cref{fig:graff}.

To reach a simple approximate formula, we consider the limit when the sources are close $k_pr_0 \rightarrow 0$, and also evaluate \eqref{eqn:src_graff} in the far field $k_pr \rightarrow \infty$. The information count should not significant change in the far-field, but it does simplify the field. Taking these asymptotic limits and retaining up too $\mathcal O(k_p^2r_0^2)$, while keeping only the leading for $k_pr \rightarrow \infty$, leads to:
 
\begin{equation}\label{eqn:asymp}
    \phi_{\text{src}} = \frac{1 + \im}{8 \sqrt{\pi k_pr}}\e^{\im k_p r}\left[ 2(A+B) + 2\im (A-B)k_pr_0 \cos\theta - (A+B) (k_pr_0\cos\theta)^2 \right].
\end{equation}
Now to distinguish between the sources we need to determine $r_0$ by measuring the above. We can further simplify this by specialising to the case where the boundary data is smooth, which will lead to a lower bound on what can be measured. In this case, we consider that $B$ is a smooth function of $r_0$ such that asymptotically:
\[
B = A + \frac{\beta}{2} r_0^2,
\]
which substituted into \eqref{eqn:asymp} leads to:
\begin{equation}\label{eqn:asymp2}
    \phi_{\text{src}} = \frac{1 + \im}{8 \sqrt{\pi k_pr}}\e^{\im k_p r} \left[4 A + r_0^2 \beta - 2A (r_0 k_p \cos \theta)^2 \right].
\end{equation}
The $\beta$ term can be anything, but will change from a negative and positive value as the $\vec r^{(2)}$ changes. When $\beta$ has the same sign of $A$ it will make it easier to determine $r_0$, and when the signs are opposite it will make it harder to determine $r_0$. To reach a simple approximation we take $\beta = 0$. 
Then, to resolve the difference between the sources at $\vec r^{(1)}$ and $\vec r^{(2)}$ we need to easily measure $r_0$, which implies that the quadratic term $r_0^2$ in \eqref{eqn:asymp2} needs to be greater or equal to the leading term, that is
\begin{equation}\label{eqn:leading_ineq}
    \left|k_p r_0 \cos\theta\right|^2 \geq 2 
\end{equation}
which is guaranteed to hold if $|k_p r_0| \geq \sqrt{2}$.

\begin{figure}[!h]
    \centering
    \includegraphics[width=0.55\linewidth]{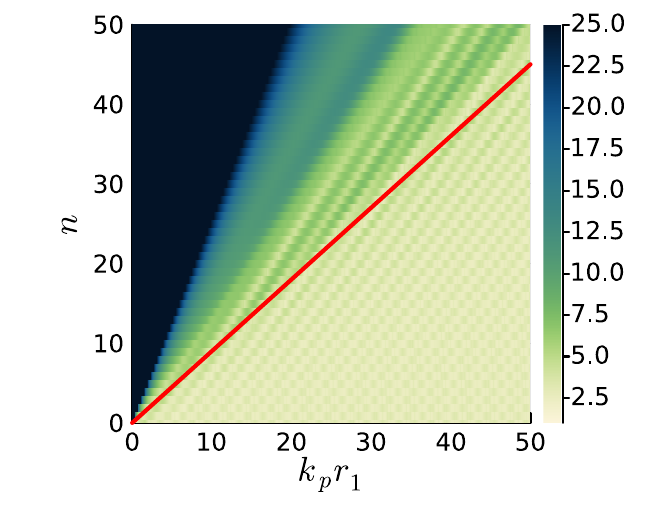}
        \caption{Figure shows the condition number plot from \Cref{fig:condition-heatmaps} with the diffraction limit in \Cref{eqn:diffraction} plotted on top in red. The diffraction limit gives a rough idea of when the inverse problem is well-posed; though it only a loose idea as we have not taken account of multiple scattering events and have performed the calculation in free space.}  
    \label{fig:diffraction}
\end{figure}

The main method of the paper uses a modal Fourier decomposition \eqref{eqn:potentials} to solve for elastic waves. For these, the level of detail of the boundary data increases with the Fourier order $n$. 
The minimal level of detail captured by order $n$ is the distance on the boundary between a trough and a crest which is equal to $\theta = r_1 |n| / \pi$. From \Cref{fig:graff} we see that this leads to the choice $2 \alpha = \pi / |n| $ and 
\begin{equation}\label{eqn:rzero}
    r_0 = r_1 \sin\left(\frac{\pi}{2|n|}\right) \approx \frac{\pi r_1}{2|n|},
\end{equation}
where the approximation is accurate for $|n| > 1$. Substituting the above into $|k_p r_0| \geq \sqrt{2}$
yields the following useful result
\begin{equation}\label{eqn:diffraction}
    \pi k_p r_1 \geq \sqrt{8} |n|.
\end{equation}
\Cref{fig:diffraction} shows when equality holds in the above superimposed on the condition number plot in \Cref{fig:condition-heatmaps}. It is clear that the limit loosely indicates where we begin to lose precision due to an ill-condition system. In reality, it is clear, that the true bound depends on more factors than those found in \Cref{eqn:diffraction}.

\printbibliography

\end{document}